\documentclass[11pt,a4paper]{article}
\pdfoutput=1
\usepackage[T1]{fontenc}
\usepackage[utf8]{inputenc}
\usepackage[top=2.5cm,bottom=2.5cm,left=2.5cm,right=2.5cm]{geometry}
\usepackage{amsmath,amssymb,amsfonts,mathrsfs,amsthm,mathtools}
\usepackage{graphicx,xcolor,booktabs,array,longtable,ragged2e,microtype}
\usepackage[colorlinks=true]{hyperref}
\definecolor{ABJMlink}{HTML}{176C68}
\definecolor{ABJMcite}{HTML}{8A4F6B}
\hypersetup{linkcolor=ABJMlink,urlcolor=ABJMlink,citecolor=ABJMcite,
 pdftitle={Modular Structure of the ADHM Effective Twisted Superpotential},
 pdfauthor={Seyed Morteza Hosseini}}
\usepackage{caption,orcidlink,cite,enumitem}
\graphicspath{{figures/}}
\theoremstyle{plain}
\newtheorem{theorem}{Theorem}[section]
\newtheorem{proposition}[theorem]{Proposition}

\theoremstyle{definition}
\newtheorem{conjecture}[theorem]{Conjecture}

\numberwithin{equation}{section}
\DeclareMathOperator{\Tr}{Tr}
\DeclareMathOperator{\re}{Re}
\DeclareMathOperator{\im}{Im}
\DeclareMathOperator{\Li}{Li}
\DeclareMathOperator{\sech}{sech}
\newcommand{\ii}{\mathrm{i}}
\newcommand{\e}{\mathrm{e}}
\newcommand{\rd}{\mathrm{d}}
\newcommand{\cW}{\mathcal{W}}
\newcommand{\W}{\mathscr{W}}
\newcommand{\Wt}{\widetilde{\mathcal{W}}}
\newcommand{\cL}{\mathcal{L}}

\newcommand{\Nh}{\widehat N}
\newcommand{\lamh}{\widehat\lambda}
\newcommand{\Cltwo}{\mathrm{Cl}_2}
\newcommand{\U}{\mathrm U}
\newcommand{\be}{\begin{equation}}
\newcommand{\ee}{\end{equation}}
\newcommand{\bea}{\begin{equation}\begin{aligned}}
\newcommand{\eea}{\end{aligned}\end{equation}}

\title{\vspace{-1.5cm}\bfseries
 Modular Structure of the ADHM\\
 Effective Twisted Superpotential}
\author{Seyed Morteza Hosseini\,\orcidlink{0000-0001-8205-400X}\\[0.5em]
 \small Centre for Theoretical Physics, Department of Physics and Astronomy,\\
 \small Queen Mary University of London, London E1 4NS, United Kingdom\\[0.5em]
}
\date{\today}
\begin{document}
\maketitle
\thispagestyle{empty}
\begin{abstract}
The Cardy limit of M2-brane indices retains nonperturbative information through the effective twisted superpotential, including corrections exponentially suppressed at large rank. We investigate the flavor dependence of these corrections in the three-dimensional $\mathcal N=4$ $\mathrm U(N)$ ADHM theory at the universal twist. For $N_f=1,2,4$ fundamental hypermultiplets, we conjecture that the full exponential remainder of the selected Bethe critical value is an absolutely convergent divisor sum obtained by applying a first-order differential operator to an Eichler integral of a weight-four Eisenstein form. Resolving this remainder requires the additive constant, which we evaluate from its known integral representation as finite Clausen sums for every positive integer $N_f$. Expanding the integral together with the shifted-rank power determines the type-IIA coefficients of the perturbative contribution at arbitrary genus and their factorial growth. With this subtraction fixed, the proposed sums reproduce independently computed Bethe critical values at up to $800$-digit accuracy, including low ranks that test the continuation beyond the reconstructed sectors. The flavor-one expression agrees with level-one ABJM, whereas the two- and four-flavor completions have distinct divisor laws related by exact dissection identities to finite combinations of ABJM functions. The comparison supports a shared weight-four organization of nonperturbative Bethe data beyond the mirror pair, with arithmetic that distinguishes the theories. Through the finite-rank Cardy relation, the proposed completion predicts the entire $1/\omega$ coefficient of the selected vacuum contribution to the superconformal index, retaining these theory-dependent corrections in its leading exponent.
\end{abstract}
\clearpage
\setcounter{page}{1}
\noindent\rule{\textwidth}{0.6pt}\vspace{-0.5cm}
\begingroup\hypersetup{linktoc=page}{\footnotesize\tableofcontents}\endgroup
\vspace{0.2cm}\noindent\rule{\textwidth}{0.6pt}
\clearpage

\section{Introduction}
\label{sec:intro}

At fixed Chern--Simons level or flavor number, protected observables of M2-brane theories admit large-$N$
expansions whose algebraic corrections can often be resummed into compact
expressions~\cite{Marino:2011eh,Bobev:2023lkx}. Their finite-rank values also contain exponentially small terms
that are invisible to any fixed order in $1/N$. The coefficients of these terms carry information beyond the
common $N^{3/2}$ scaling and its perturbative corrections. For supersymmetric indices that reproduce the
leading entropy of $\mathrm{AdS}_4$ black holes~\cite{Benini:2015eyy}, determining this information supplies
field-theory data beyond the classical gravitational limit. A nonperturbative description must account for
both the exponential scales and the relations among their coefficients.

The effective twisted superpotential $\Wt$ determines the Bethe
vacua of three-dimensional $\mathcal N=2$ gauge
theories~\cite{Nekrasov:2009uh}. Let $\mathcal I_{\mathrm{vac}}$
denote the contribution of a selected Bethe vacuum to the
superconformal index
$\mathcal I$~\cite{Bhattacharya:2008bja,Bhattacharya:2008zy,Kim:2009wb},
and $\Wt_{\mathrm{vac}}$ the corresponding critical value.
We parametrize the angular fugacity by $x=\e^{\ii\pi\omega}$.
Taking the Cardy limit $\omega\to\ii0^+$ in the localization
formula at fixed rank and chemical potentials, and evaluating
the selected saddle, gives
\begin{equation}
 \log\mathcal I_{\mathrm{vac}}(\omega)
 =-\frac{\operatorname{Im}\Wt_{\mathrm{vac}}}{\pi\omega}
   +O(\omega^0)\,,
 \label{eq:intro-cardy}
\end{equation}
in the normalization used
below~\cite{Choi:2019dfu,Bobev:2022wem,Bobev:2024mqw}.
This relation retains the full rank dependence of the critical
value, so corrections exponentially small in $N$ contribute
to the leading $1/\omega$ coefficient. The critical values
also govern holomorphic-block
asymptotics~\cite{Beem:2012mb}, while the Hessian and effective
dilaton enter the additional factors required for twisted
partition functions~\cite{Benini:2015noa,Closset:2018ghr}.
For holographic M2-brane theories, the leading large-$N$
critical value supplies the field-theory input to gravitational
blocks whose gluing yields supersymmetric $\mathrm{AdS}_4$
black hole entropy functions~\cite{Hosseini:2019iad}.

The sphere partition function gives an instructive example of how a compact perturbative expression can be
extended nonperturbatively. The ABJM Fermi gas resums the perturbative sphere partition function into an Airy
function and resolves its constant map and instanton sectors~\cite{Marino:2011eh,Hatsuda:2013oxa}, and its
special-level grand partition functions admit theta-function representations tied to a spectral
curve~\cite{Codesido:2014oua}. More general Airy formulas have been proposed for mass-deformed,
squashed-sphere observables~\cite{Bobev:2025ltz}. Recent factorization relations compare the superconformal~\cite{Bhattacharya:2008bja,Bhattacharya:2008zy,Kim:2009wb} and
twisted indices with the squashed-sphere partition function through matched Cardy and large-squashing saddle
expansions~\cite{Bobev:2026lvl}. For ADHM with one fundamental hypermultiplet, the holomorphic blocks also admit
geometric descriptions in terms of vortex moduli spaces and equivariant quantum
K-theory~\cite{Crew:2020psc}. These constructions constrain several realizations of protected data. A
spectral or geometric characterization that fixes the exponential sectors of the finite-rank Bethe critical
value remains to be derived.

Recent ABJM calculations suggest an arithmetic organization of this critical
value~\cite{Hosseini:2026dkj,Hosseini:2026lvj}. At the universal twist and levels $k=1,2,4$, the reconstructed exponential
coefficients are consistent with cubic divisor laws. Their conjectural all-order sums are obtained from
Eichler integrals of weight-four Eisenstein forms. The relation between divisor sums, Lambert series and
Eisenstein primitives constrains the continuation once the relevant Eisenstein subspace is
specified~\cite{Dorigoni:2020oon}. The physical assertion is that the solution of the Bethe equations
selects this subspace. We ask whether the same weight-four organization governs the critical twisted
superpotential of a different M2-brane theory, and how its arithmetic depends on the matter content.

We study the $\mathcal N=4$ ADHM theory with gauge group $\U(N)$, one adjoint hypermultiplet and $N_f$
fundamental hypermultiplets. Its M2-brane geometry is $\mathbb C^2\times\mathbb C^2/\mathbb Z_{N_f}$, whereas
ABJM has the diagonal quotient $\mathbb C^4/\mathbb Z_k$~\cite{Aharony:2008ug,Bobev:2023lkx}. At $N_f=1$, mirror
symmetry relates ADHM to level-one ABJM~\cite{Kapustin:2010xq}. For general flavor, mirror Chern--Simons
descriptions involve longer circular quivers with additional zero-level nodes~\cite{Hatsuda:2015lpa}. Exact
relations to ordinary two-node ABJ(M) also concern particular observables. The Coulomb-limit superconformal
indices agree at finite rank for general $N_f=k$, with the appropriate fugacity
map~\cite{Okazaki:2022sxo,Hayashi:2022ldo}. This equality counts a protected sector and is distinct
from the Cardy limit in~\eqref{eq:intro-cardy}. At $N_f=2$, the round-sphere matrix integral equals the
normalized magnitude of the $\U(N)_2\times\U(N+1)_{-2}$ ABJ partition function, with the pure $\U(1)_2$
Chern--Simons factor removed~\cite{Hatsuda:2014vsa}. For $N_f>1$, these observable identities do not by
themselves identify the full ADHM theory with equal-rank ABJM.

Modular objects already occur in these other ADHM observables. The two-flavor sphere identity transfers the
ABJ theta-function completion to its grand partition function~\cite{Codesido:2014oua}. Affine
$\widehat{\mathfrak{su}}(N_f)_1$ characters also appear in proposed leading giant-graviton coefficients of
ADHM Higgs indices, including $N_f=2,4$, with extensions to higher wrapping
number~\cite{Hayashi:2024aaf,Hayashi:2025ukc}. These character formulas identify modular building blocks within
index expansions, without establishing modularity of the full finite-rank index. The character construction
extends to other flavors and does not single out $N_f=2,4$. Its modular variable is an index fugacity, while
sphere spectral modularity involves periods of the associated curve. The question addressed here concerns a
different variable, constructed from the rank and flavor number after solving the Bethe equations, and a
different observable, the normalized critical twisted superpotential. The defining Bethe equations are
rational equations in exponentiated gauge holonomies with chemical potentials as parameters. They exhibit no
apparent modular transformation of this rank-dependent nome.

For ADHM, the continuum Bethe solution determines the leading critical value and a leading large-$N$ relation
to ABJM~\cite{Hosseini:2016ume}. Subsequent calculations reconstructed the algebraic corrections and isolated
the constant and exponential remainder~\cite{Bobev:2023lkx,Bobev:2022wem}. We work at the universal specialization
$\Delta_1=\Delta_2=\Delta_q=\Delta_{\widetilde q}=\pi/2$,
$\Delta_3=\pi$ and $\Delta_m=0$. On the selected branch, we set
$\W_{\mathrm A}(N,N_f)=\operatorname{Im}\Wt_{\mathrm{ADHM}}$. Its proposed perturbative contribution in our
normalization is
\begin{equation}
 \W_{\mathrm A}^{\mathrm{pert}}(N,N_f)
 =\frac{\pi^2}{3}\sqrt{\frac{N_f}{2}}\,\Nh^{3/2}+C(N_f)\,,
 \qquad
 \Nh=N+\frac{7N_f}{24}+\frac{1}{3N_f}\,.
 \label{eq:intro-parent}
\end{equation}
The identification of this shifted power as the complete algebraic contribution is supported by finite-rank
numerics. A convergent integral for $C(N_f)$ and its relation to ABJM constants are already known~\cite{Hong:2026zul}.
An explicit evaluation fixes the subtraction needed to resolve the remainder, since any error in the additive
constant eventually exceeds every exponentially small sector.

For $N_f=1,2,4$, we conjecture the finite-rank completion
\begin{equation}
 \begin{aligned}
 t&=2\pi\sqrt{\frac{2\Nh}{N_f}}\,,\qquad q=\e^{-t}\,,\\
 \W_{\mathrm A}(N,N_f)
 &=\W_{\mathrm A}^{\mathrm{pert}}(N,N_f)
   +\frac{N_f^2}{2\pi}\sum_{m\geq1}c_m^{(N_f)}
       \left(t+\frac1m\right)q^m\,.
 \end{aligned}
 \label{eq:intro-completion}
\end{equation}
The nome $q$ is distinct from the angular fugacity $x$ and remains fixed in the Cardy limit. The quantities
$m^2c_m^{(N_f)}$ are finite rational combinations of the cubic divisor sum $\sigma_3(m)=\sum_{d\mid m}d^3$ and
its dilations, with $\sigma_3(x)=0$ when $x$ is not a positive integer. The flavor-one expression agrees with
the inherited level-one ABJM completion after matching the branch and additive normalization. The flavor-two
and flavor-four sequences are reconstructed independently from ADHM data. Their first coefficients are
$c_1^{(2)}=-1$ and $c_1^{(4)}=1/2$, compared with $c_1^{(1)}=5$. In particular, the two-flavor correction has
the opposite sign. The higher sectors resolve further divisibility patterns that cannot be obtained by
replacing the ABJM level with the flavor number.

With $D=q\partial_q$ acting at fixed $t$, the proposed correction has the form $(1+tD)\Phi_{N_f}(q)$,
evaluated at $q=\e^{-t}$. The function $\Phi_{N_f}$ is the zero-constant Fourier primitive of a
constant-subtracted weight-four Eisenstein combination, with three inverse powers of $D$. In the physical
coordinate $q=\e^{2\pi\ii\tau}$, the generating forms lie in $M_4(\Gamma_0(L))$ with minimal levels $L=4,4,8$
for $N_f=1,2,4$. In the shifted coordinate $\tau+1/2$, their minimal levels are $2,4,8$. We prove these level
statements, absolute convergence and explicit truncation bounds. The power-series components have radius of
convergence one, so the proposed correction is well defined at every positive physical rank. These are
statements about the generators and their Eichler primitives, not a homogeneous modular transformation law for
$\W_{\mathrm A}$. Lambert-series dissections express the proposed ADHM functions as finite combinations of the
special-level ABJM functions at rescaled actions. At $N_f=2,4$, no single affine change of rank, overall
scaling and additive constant identifies them with an ABJM expression in the tested class $k=1,2,4$. The
finite combinations generally involve noninteger rank arguments and do not establish new gauge-theory
dualities. The resulting identities relate functions with a common modular construction and different
arithmetic.

The comparison is made on a specified finite-rank branch.%
\footnote{At $N_f=1$, additional solutions of the Bethe equations and
their contributions to supersymmetric indices have been studied
in \cite{Peng:2026udd}.}
A strictly convex real potential proves existence
and uniqueness of the ordered imaginary solution of the logarithmic Bethe equations, together with reflection
symmetry. Starting from the known constant-map integral and its two-level ABJM
relation~\cite{Hong:2026zul,Hosseini:2026lvj}, we derive finite Clausen evaluations for every positive integer $N_f$ and
fix the remaining normalization analytically. Combining the known large-flavor expansion of the integral with
the shifted power gives closed coefficients at arbitrary genus and their factorial growth. This is the
type-IIA expansion of~\eqref{eq:intro-parent}. It does not give the complete physical genus coefficients at
fixed $\lambda=N/N_f$. Indeed, the nome tends to $\exp[-2\pi\sqrt{2(\lambda+7/24)}]\neq0$ as $N_f\to\infty$,
so fixed-flavor exponential sectors can contribute at fixed coupling. Incorporating them requires a
generic-flavor remainder with control uniform in this limit.

High-precision Bethe solutions test the proposed completions at $897$ special-flavor points, with $2\leq
N\leq300$. Root and critical-value enclosures separate numerical uncertainty from series truncation, with
$800$-digit targets through $N=100$. At flavors two and four, the relative coefficient of $t$ and the constant
in each sector is fitted without imposing the relation in~\eqref{eq:intro-completion}. Comparisons of the full
convergent sums include low ranks outside the fitting windows. There, reaching the precision of the critical
values requires many more than the sixteen fitted sectors. Agreement tests the proposed functions beyond the
coefficient range used in their reconstruction. They do not independently determine every coefficient or prove
the all-order identification with the Bethe critical value.

Assuming that identification, \eqref{eq:intro-completion} fixes the entire $1/\omega$ coefficient of the
selected contribution in~\eqref{eq:intro-cardy}, including its exponentially small finite-$N$ terms. It fixes
the same coefficient for the full superconformal index in a chamber where this contribution dominates.
Dominance depends on the integration cycle and competing saddles, as also emphasized in recent analyses of the
Cardy limit~\cite{ArabiArdehali:2025bub}, and does not follow from uniqueness on the real logarithmic branch.
A derivation of the completion from the Bethe equations would have to explain why their critical value belongs
to the proposed Eisenstein class. The two- and four-flavor results give concrete coefficient laws against
which such a derivation can be tested.

Section~\ref{sec:branch} defines the Bethe problem and its branch, and Section~\ref{sec:constant} evaluates
the constant. Sections~\ref{sec:reconstruction}--\ref{sec:relations} develop the exponential reconstruction,
modular analysis, type-IIA expansion and ABJM relations. The numerical tests and discussion, including
the Cardy application, are given in Sections~\ref{sec:validation} and~\ref{sec:discussion}.
The appendices give a root-of-unity representation of the constant, the residual-to-root enclosure and
details of the numerical implementation.

\section{The ADHM Bethe problem}\label{sec:branch}
\subsection{Matter content and the universal specialization}
The ADHM theory is the three-dimensional $\mathcal N=4$ gauge theory
with gauge group $\U(N)$, vanishing Chern--Simons level, and $N_f\in\mathbb Z_{>0}$
fundamental hypermultiplets. In $\mathcal N=2$ notation it contains three
adjoint chiral multiplets $\Psi_I$, $I=1,2,3$, and fundamental--antifundamental
pairs $\psi_q,\widetilde\psi_q$, $q=1,\ldots,N_f$. Their ordinary matter
superpotential is
\be
 W_{\rm matter}=\Tr\!\left[
 \sum_{q=1}^{N_f}\widetilde\psi_q\Psi_3\psi_q
 +\Psi_3[\Psi_1,\Psi_2]\right]\,.
 \label{eq:matter_superpotential}
\ee
We follow the Bethe-potential conventions of
Refs.~\cite{Bobev:2023lkx,Hong:2026zul}, converting the chemical potentials to
angular variables by $\Delta_{\rm there}=\Delta/\pi$.
All holonomies, chemical potentials and twisted superpotentials below
are dimensionless. The matter interactions impose
\be
 \Delta_1+\Delta_2+\Delta_3=2\pi\,,\qquad
 \Delta_q+\Delta_{\widetilde q}+\Delta_3=2\pi\,.
 \label{eq:chemical_constraints}
\ee
The topological chemical potential is denoted by $\Delta_m$.
The universal specialization is
\be
 \Delta_1=\Delta_2=\Delta_q=\Delta_{\widetilde q}=\frac\pi2\,,
 \qquad \Delta_3=\pi\,,\qquad\Delta_m=0\,.
 \label{eq:universal}
\ee
In the four-parameter notation of Ref.~\cite[Eqs.~(48)--(49)]{Hong:2026zul},
the four entries are
\[
 (\widetilde\Delta_1,\widetilde\Delta_2,\widetilde\Delta_3,\widetilde\Delta_4)
 =(\Delta_1,\Delta_2,\Delta_3/2-\Delta_m/N_f,\Delta_3/2+\Delta_m/N_f)\,,
\]
with $\widetilde\Delta_a=\pi/2$ for every $a$ at the universal specialization.
The equality of the two fundamental chemical potentials fixes a symmetric
representative of their constrained sum. Magnetic fluxes specify the
associated twist, but do not enter the Bethe equations or the critical
twisted superpotential evaluated here.

Throughout, $\ii=\sqrt{-1}$ with positive imaginary part, $\log$ is the
natural logarithm, and $\log_{10}$ is the base-ten logarithm.
We use the principal polylogarithms
$\Li_s(z)=\sum_{r\geq1}z^r/r^s$ for $|z|<1$, continued with their
standard cut along $[1,\infty)$.

\subsection{Exponentiated equations and the logarithmic branch}
Let $u_j$ be the complex gauge holonomies and
$x_j=\e^{\ii u_j}$, $j=1,\ldots,N$. We choose the logarithmic
branch integers
\be
 \ell_j=\left\lfloor\frac{N+1}{2}\right\rfloor-j\,,
 \label{eq:branch_integers}
\ee
and assign each chiral weight the
contribution $\Li_2(\e^{\ii z})-\tfrac12g_2(z)$, where
$g_2(z)=z^2/2-\pi z+\pi^2/3$ and $z$ includes the gauge
and flavor arguments. In the strip $0<\re z<2\pi$, the
inversion identity
$\Li_2(\e^{\ii z})+\Li_2(\e^{-\ii z})=g_2(z)$
expresses this contribution as an antisymmetric combination
of dilogarithms. In our angular conventions, the ADHM
twisted superpotential of Ref.~\cite[Eq.~(47)]{Hong:2026zul},
with its holonomy-independent term adjusted to this
normalization, becomes
\bea
 \Wt(\boldsymbol u;\Delta)={}&
 \sum_{j=1}^N
 \left[\pi(N+1-2j)-\Delta_m
       -\frac{N_f}{2}(2\pi-\Delta_q-\Delta_{\widetilde q})
 \right]u_j
 \\
 &+\frac12\sum_{I=1}^3\sum_{j,h=1}^N
 \left[
 \Li_2\!\left(\e^{\ii(u_h-u_j+\Delta_I)}\right)
 -\Li_2\!\left(\e^{\ii(u_h-u_j-\Delta_I)}\right)
 \right]
 \\
 &+N_f\sum_{j=1}^N
 \left[
 \Li_2\!\left(\e^{\ii(-u_j+\Delta_{\widetilde q})}\right)
 -\Li_2\!\left(\e^{\ii(-u_j-\Delta_q)}\right)
 \right]
 \\
 &+\frac{NN_f}{2}
 \left[g_2(\Delta_q)-g_2(\Delta_{\widetilde q})\right]\,.
 \label{eq:twisted_superpotential}
\eea
The diagonal adjoint terms are retained because they
contribute to the critical value, although they are
independent of the holonomies. The final term implements
the same normalization for the fundamental chirals and
vanishes when $\Delta_q=\Delta_{\widetilde q}$.

At the universal specialization~\eqref{eq:universal},
the Bethe equations
$\exp(\ii\partial\Wt/\partial u_j)=1$ reduce to
\be
 \left(\frac{1+\ii x_j}{x_j+\ii}\right)^{N_f}
 \prod_{h\ne j}
 \left(\frac{x_j-\ii x_h}{x_j+\ii x_h}\right)^2=1\,.
 \label{eq:exponentiated}
\ee
The adjoint contribution at $\Delta_3=\pi$ cancels, while
the two adjoints at $\Delta_1=\Delta_2=\pi/2$ give the
squared pair factor. Only off-diagonal weights enter
this product, since the diagonal terms disappear upon
differentiation. Inserting a formal self-factor would
multiply the left-hand side by $(-\ii)^2=-1$ and change
the equation.

We consider imaginary holonomies $u_j=\ii y_j$ on the
branch~\eqref{eq:branch_integers}. Define the smooth odd
function
\be
 f(y)=\arctan(\sinh y)
     =\frac\pi2-2\arctan(\e^{-y})\,,
 \qquad
 f'(y)=\sech y>0\,,
 \label{eq:f_definition}
\ee
where the inverse tangent takes its real principal value.
The resulting unwrapped equations are
\be
 B_j(\boldsymbol y)
 =N_f f(y_j)+2\sum_{h=1}^N f(y_j-y_h)
       -\pi(2j-N-1)=0\,,
 \qquad j=1,\ldots,N\,.
 \label{eq:bethe}
\ee
The term $h=j$ in this sum vanishes because $f(0)=0$.

For $x_j=\e^{-y_j}>0$, the fundamental ratio in
Eq.~\eqref{eq:exponentiated} is $\e^{-\ii f(y_j)}$,
and each squared pair ratio is
$-\e^{-2\ii f(y_j-y_h)}$. Its left-hand side is therefore
\be
 (-1)^{N-1}
 \exp\!\left[
 -\ii N_f f(y_j)-2\ii\sum_{h\ne j}f(y_j-y_h)
 \right]
 =\e^{-\ii B_j(\boldsymbol y)}\,.
\ee
Thus Eq.~\eqref{eq:bethe} implies the exponentiated
equations with the required signs. Exponentiation retains
only the conditions $B_j\in2\pi\mathbb Z$, so it does not
distinguish the chosen branch from other logarithmic
solutions. Both the numerical solve and its independent
verification impose $B_j=0$ as specified in
Eq.~\eqref{eq:bethe}.

\subsection{Existence, ordering and uniqueness}
The real logarithmic problem has a variational formulation that fixes
the branch independently of a numerical initial guess. Define the Clausen
function and Catalan's constant by
\be
 \Cltwo(\theta)=\sum_{r=1}^{\infty}\frac{\sin(r\theta)}{r^2}
 =\im\Li_2(\e^{\ii\theta})\,,\qquad G=\Cltwo(\pi/2)\,.
 \label{eq:Clausen}
\ee
The series is absolutely convergent for real $\theta$. An even primitive
of $f$ is
\be
 H(y)=\int_0^y f(s)\,\rd s
 =\frac\pi2|y|+2\im\Li_2(\ii\e^{-|y|})-2G\,.
 \label{sm:primitive}
\ee
For $y>0$, differentiating the right-hand side gives
$\pi/2-2\arctan(\e^{-y})$; evenness and its value at zero complete
the verification on the real line. The dilogarithm term is nonnegative,
as follows from
$\im\Li_2(\ii r)=\int_0^r\arctan s\,\rd s/s$ for $r\geq0$.

Introduce the real variational function
\be
 \Psi(\boldsymbol y)=N_f\sum_jH(y_j)
 +2\sum_{i<j}H(y_i-y_j)
 -\pi\sum_j(2j-N-1)y_j\,.
 \label{sm:Psi}
\ee
Its gradient is $\boldsymbol B$. The following statement refers to
this real logarithmic problem, allowing real $N_f>0$ as a mathematical
continuation of the physical integer flavor number.

\begin{proposition}[The ordered imaginary branch]\label{prop:branch}
For every integer $N\geq1$ and real $N_f>0$, Eq.~\eqref{eq:bethe}
has exactly one real solution $\boldsymbol y^*$. It satisfies
\be
 y_1^*<\cdots<y_N^*\,,\qquad
 y_j^*=-y_{N+1-j}^*\,.
 \label{eq:ordered_branch}
\ee
In particular, the middle root vanishes exactly when $N$ is odd.
\end{proposition}
\begin{proof}
For every nonzero real vector $\boldsymbol z$, the Hessian obeys
\be
 \boldsymbol z^{\mathsf T}(\nabla^2\Psi)\boldsymbol z
 =N_f\sum_j\sech y_j\,z_j^2
 +2\sum_{i<j}\sech(y_i-y_j)(z_i-z_j)^2>0\,.
 \label{sm:convexity}
\ee
Thus $\Psi$ is strictly convex. Equation~\eqref{sm:primitive} and
\be
 \sum_j(2j-N-1)y_j
 =\sum_{i<j}(y_j-y_i)\leq\sum_{i<j}|y_j-y_i|
\ee
also imply
\be
 \Psi(\boldsymbol y)\geq\frac{\pi N_f}{2}\sum_j|y_j|
 -2GN_fN-4G\binom N2\,.
 \label{sm:coercivity}
\ee
Consequently $\Psi$ is coercive and attains a minimum on
$\mathbb R^N$. Strict convexity makes that minimum its unique critical
point.

If $i<j$ but $y_i\geq y_j$ at a zero of $\boldsymbol B$, then
\be
 B_i-B_j=N_f[f(y_i)-f(y_j)]
 +2\sum_h[f(y_i-y_h)-f(y_j-y_h)]+2\pi(j-i)>0\,.
\ee
Monotonicity of $f$ gives the strict inequality, contradicting
$B_i=B_j=0$. This proves ordering. Oddness of $f$ shows that
$y_j\mapsto-y_{N+1-j}$ maps a solution to a solution, so uniqueness
imposes reflection symmetry.
\end{proof}

Neighboring-rank continuation supplies efficient initial guesses for
this solution. Once a root is enclosed for Eq.~\eqref{eq:bethe}, the
proposition excludes a jump to another solution of the same real
logarithmic problem. It does not classify other complex Bethe vacua or
determine which terms dominate a complete vacuum sum.

\subsection{The critical value and its additive normalization}
At the universal specialization, Eq.~\eqref{eq:twisted_superpotential}
restricts on the imaginary locus to
\bea
 \Wt(\ii\boldsymbol y)={}&
 \ii\pi\sum_{j=1}^N\left(N+1-2j-\frac{N_f}{2}\right)y_j
 +\sum_{j,h=1}^N\left[\Li_2(\ii\e^{y_j-y_h})
                       -\Li_2(-\ii\e^{y_j-y_h})\right]
 \\
 &+N_f\sum_{j=1}^N\left[\Li_2(\ii\e^{y_j})
                       -\Li_2(-\ii\e^{y_j})\right]\,.
 \label{eq:offshell}
\eea
The double sum includes diagonal terms. This convention fixes the
additive constant, which is essential when comparing rank-independent
contributions. Principal continuation along the imaginary axis gives
$\Li_2(-\ii r)=\overline{\Li_2(\ii r)}$ for real $r>0$.
Every dilogarithm difference in Eq.~\eqref{eq:offshell}, as well as its
linear term, is therefore purely imaginary for every real input
$\boldsymbol y$. Define
\be
 \W_{\mathrm A}(N,N_f)=\im\Wt(\ii\boldsymbol y^*)\,,\qquad
 \Wt(\ii\boldsymbol y^*)=\ii\W_{\mathrm A}(N,N_f)\,.
 \label{eq:observables}
\ee
The zero real part is an analytic consequence of this representative,
not a separate numerical observation at the precision floor.

For $a\geq0$, differentiating and setting $a=0$ proves
\be
 \im\Li_2(\ii\e^a)-\im\Li_2(\ii\e^{-a})=\frac\pi2a\,.
\ee
Use this identity to pair the two ordered adjoint terms at every
$i<j$. Their linear pieces cancel the first term in
Eq.~\eqref{eq:offshell} by
$\sum_{i<j}(y_j-y_i)=\sum_j(2j-N-1)y_j$.
The remaining fundamental linear contribution combines with its
negative-argument counterpart to give an absolute value. Hence, on
ordered inputs,
\bea
 \im\Wt(\ii\boldsymbol y)={}&
 2NG+4\sum_{i<j}\im\Li_2(\ii\e^{-(y_j-y_i)})
 +N_f\sum_j\left[2\im\Li_2(\ii\e^{-|y_j|})
                              +\frac\pi2|y_j|\right]
 \\
 ={}&\Psi(\boldsymbol y)+2G(N^2+N_fN)\,.
 \label{sm:potential}
\eea
The decaying arguments make this expression suitable for high-precision
evaluation. At $N=1$, the unique root is $y_1^*=0$, giving the exact
benchmark
\be
 \W_{\mathrm A}(1,N_f)=2(N_f+1)G\,.
 \label{eq:N1_benchmark}
\ee

For a different real fundamental representative with
$\Delta_q+\Delta_{\widetilde q}=\pi$ and $\Delta_m=0$, set
$v_j=u_j+\Delta_q-\pi/2$. The dilogarithm arguments become those of
the symmetric representative, while the holonomy-independent term
cancels the change in the linear term. Therefore
\be
 \Wt_{\Delta_q}(\boldsymbol u)
 =\Wt_{\mathrm{sym}}(\boldsymbol v)\,.
 \label{eq:representative_shift}
\ee
The full twisted superpotential is unchanged by this change of
representative and holonomy coordinate. All numerical values in
this work use Eq.~\eqref{eq:universal} and the normalization of
Eq.~\eqref{eq:offshell}.

\subsection{The continuum density and the leading large-rank term}
At fixed $N_f$, the continuum Bethe solution gives a useful normalization
check and explains the scale of the selected roots
\cite[Eqs.~(3.4)--(3.11)]{Hosseini:2016ume}. Introduce the coordinate
$\xi=y/\sqrt N$ and a nonnegative density $\rho(\xi)$ normalized by
$\int\rho(\xi)\,\rd\xi=1$. In the continuum approximation, sums over
the roots are replaced by this density, and the short-range adjoint
interaction in Eq.~\eqref{sm:potential} becomes local. Its integrated
kernel is
\be
 \int_0^\infty\im\Li_2(\ii\e^{-s})\,\rd s
 =\sum_{r=0}^\infty\frac{(-1)^r}{(2r+1)^3}
 =\frac{\pi^3}{32}\,.
 \label{eq:continuum_kernel}
\ee
The first equality follows by termwise integration of an absolutely
integrable series. The last follows by integrating
$\sum_{m\geq1}\cos(m\theta)/m^2=\pi^2/6-\pi\theta/2+\theta^2/4$
from $\theta=0$ to $\pi/2$. Each root contributes interactions within a finite distance in
$y$, corresponding to a distance $O(N^{-1/2})$ in $\xi$.
Replacing the neighboring density by its local value therefore gives
\be
 \frac{\W_{\mathrm A}}{N^{3/2}}
 \sim\frac{\pi^3}{8}\int\rho(\xi)^2\,\rd\xi
       +\frac{\pi N_f}{2}\int|\xi|\rho(\xi)\,\rd\xi\,.
 \label{eq:continuum_functional}
\ee
The second term comes directly from the absolute-value fundamental
term. This reduction assumes a continuum density with the stated root
scaling; it controls the leading term and does not retain finite-spacing
corrections.

Varying Eq.~\eqref{eq:continuum_functional} at fixed normalization gives
$\pi^3\rho(\xi)/4+\pi N_f|\xi|/2=\mu$ on the support, where
$\mu$ is the Lagrange multiplier. Nonnegativity and normalization fix
\be
 \rho(\xi)=\frac{\sqrt{2N_f}}{\pi}
              -\frac{2N_f}{\pi^2}|\xi|\,,
 \qquad |\xi|\leq\frac{\pi}{\sqrt{2N_f}}\,,
 \label{eq:continuum_density}
\ee
with $\rho=0$ outside this interval. For a normalized triangular
profile on $[-L,L]$, one has
$\int\rho^2\,\rd\xi=2/(3L)$ and
$\int|\xi|\rho\,\rd\xi=L/3$.
Using $L=\pi/\sqrt{2N_f}$ in Eq.~\eqref{eq:continuum_functional}
then yields
\be
 \W_{\mathrm A}(N,N_f)
 \sim\frac{\pi^2\sqrt{2N_f}}6N^{3/2}\,.
 \label{eq:leading_continuum}
\ee
The coefficient agrees with the universal specialization of the
large-$N$ Bethe potential. The shifted rank, constant term and
exponentially small sectors require information beyond this continuum
approximation.

\section{The shifted-rank expansion and its constant map}\label{sec:constant}
\subsection{The perturbative contribution}
The large-rank analysis of the ADHM Bethe potential organizes its
algebraic corrections into a shifted $N^{3/2}$ power
\cite{Bobev:2023lkx,Hong:2026zul}. At the universal specialization, define
\be
 \Nh=N+\frac{7N_f}{24}+\frac{1}{3N_f}\,,\qquad
 t=2\pi\sqrt{\frac{2\Nh}{N_f}}\,,\qquad q=\e^{-t}\,.
 \label{eq:shift}
\ee
The positive action $t$ and the physical nome $q$ will also be used
for the exponentially small corrections. The proposed perturbative
contribution is
\be
 \W_{\mathrm A}^{\mathrm{pert}}(N,N_f)
 =\frac{\pi^2\sqrt{2N_f}}6\Nh^{3/2}+C(N_f)
 =\frac{N_f^2t^3}{96\pi}+C(N_f)\,.
 \label{eq:perturbative}
\ee
Its identification as the complete algebraic large-$N$ contribution to
Eq.~\eqref{eq:observables} remains a numerically supported asymptotic
statement. The finite-rank branch theorem does not imply this expansion.
In the conventions of Ref.~\cite[Eqs.~(50)--(54)]{Hong:2026zul},
$\W_{\mathrm A}/(2\pi)=\im \mathcal V^{\mathrm{ADHM}}/(2\pi)$ and
\be
 C(N_f)=2\pi\widehat g_0^{\mathrm{ADHM}}(N_f,\widetilde\Delta_{\rm sc})\,.
 \label{eq:constant_dictionary}
\ee
Thus both the rank shift and the integral representation used below
specialize previously proposed large-rank results. The explicit integral evaluation fixes the subtraction needed to resolve the exponential sectors.

For real $N_f>0$, take the analytic constant map to be
\begin{align}
 C(N_f)={}&2\pi a_2N_f^2-\frac\pi4\log2
 +\frac{2N_f}{\pi}\int_0^\infty\log(1-\e^{-N_fx})
 \left[\frac12\log\cosh x+\frac14\log\cosh\frac x2\right]\rd x\,,
 \label{eq:constant}\\
 a_2={}&\frac{3\zeta(3)}{32\pi^2}+\frac7{32}\log2-\frac12A(1/2)\,.
 \label{eq:a2definition}
\end{align}
Here $\zeta$ is the Riemann zeta function, and the function $A(k)$ is
normalized as the ABJM sphere constant map
\cite[Eq.~(3.13)]{Hatsuda:2014vsa},
\be
 A(k)=\frac{2\zeta(3)}{\pi^2k}\left(1-\frac{k^3}{16}\right)
 +\frac{k^2}{\pi^2}\int_0^\infty
       \frac{x\log(1-\e^{-2x})}{\e^{kx}-1}\,\rd x\,,
 \qquad k>0\,.
 \label{sm:a2}
\ee
The integral in Eq.~\eqref{eq:constant} converges absolutely. Its
integrand is $O(x^2\log x)$ at zero and decays exponentially times a
linear function at infinity. The appearance of $A$ concerns an identity
between constant-map functions and does not identify the full sphere
partition function with a Bethe observable.

\begin{figure}[tbp]
 \centering\includegraphics[width=.68\textwidth]{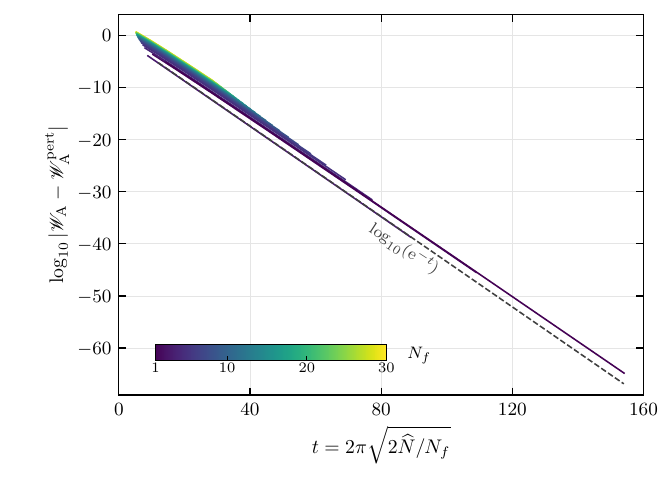}
 \caption{The logarithmic remainder $\log_{10}|\W_{\mathrm A}-\W_{\mathrm A}^{\mathrm{pert}}|$
 on the full grid $2\leq N\leq300$, $1\leq N_f\leq30$, plotted against
 the action in Eq.~\eqref{eq:shift}. The finite constant formulas below
 are used without fitting. The dashed reference has slope $-\log_{10}\e$, corresponding to an $\e^{-t}$ decay. The data support exponential suppression after
 subtracting the shifted power and constant. They do not, by themselves,
 prove that every algebraic correction is contained in the shifted
 power.}
 \label{fig:perturbative}
\end{figure}

Figure~\ref{fig:perturbative} shows the remainder after subtracting
Eq.~\eqref{eq:perturbative} from all $8,970$ scalar critical values.
No fit parameter enters the subtraction. The approximately linear decay
on the logarithmic plot, when expressed in the action $t$, supports an
exponentially small remainder. The larger flavor values require higher
ranks to separate this remainder from the algebraic expansion. Independent
fits of the shift and their rank-window dependence are discussed in the
numerical analysis.

\subsection{Reduction to two ABJM levels}
The ABJM constant map in the same normalization is
\bea
 C_{\mathrm B}(k)
 ={}&-\frac{k^2\zeta(3)}{16\pi}-\frac\pi3\log2
 +\frac{2k}{\pi}\int_0^\infty
       \log\cosh x\log(1-\e^{-kx})\,\rd x
 \\
 ={}&-\frac{k^2\zeta(3)}{16\pi}
 +\frac{2k}{\pi}\int_0^\infty
       \log(2\cosh x)\log(1-\e^{-kx})\,\rd x\,.
 \label{sm:CB}
\eea
This is $C_{\rm int}(k)$ in the analytic constant-map theorem of
Ref.~\cite[Theorem~3.1]{Hosseini:2026lvj}, and
$2\pi\widehat g_0(k,\Delta_{\rm sc})$ in
Ref.~\cite[Eq.~(29)]{Hong:2026zul}. The two displayed representations agree
because $\int_0^\infty\log(1-\e^{-kx})\,\rd x=-\pi^2/(6k)$.

The first kernel in Eq.~\eqref{eq:constant} is one half of the
integral term in $C_{\mathrm B}(N_f)$. Substitution $x=2y$ in the
second makes it one quarter of the corresponding term in
$C_{\mathrm B}(2N_f)$. The explicit logarithms then cancel, giving
\be
 C(N_f)=\frac12C_{\mathrm B}(N_f)+\frac14C_{\mathrm B}(2N_f)
 +N_f^2\left(2\pi a_2+\frac{3\zeta(3)}{32\pi}\right)\,.
 \label{eq:constantdecomp}
\ee
This two-level relation was obtained in Ref.~\cite[Eq.~(55)]{Hong:2026zul}.
Absolute convergence justifies the changes of variables for every
real $N_f>0$. Its use here allows the exact finite ABJM formulas to
be transferred to integer flavor number once $a_2$ is evaluated.

\subsection{Analytic evaluation of the normalization}
The half-integer value $A(1/2)$ can be evaluated without imposing an
equality of physical mirror constants. For positive real $a,b$, put
\bea
 L_a(x)&=\log(1-\e^{-ax})\,,\\
 D(a,b)&=\int_0^\infty L_a(x)L_b(x)\,\rd x\,,\qquad
 J(a,b)=\int_0^\infty\frac{xL_a(x)}{\e^{bx}-1}\,\rd x\,.
 \label{eq:exchange_integrals}
\eea
These are the integral families of Ref.~\cite[Appendix~B.1]{Hong:2026zul}.
Integrating the derivative of $xL_a(x)L_b(x)$ gives
\be
 D(a,b)=-bJ(a,b)-aJ(b,a)\,,\qquad
 J(a,b)=\frac4{a^2}J(2,2b/a)\,.
 \label{eq:exchangelemma}
\ee
The boundary term is $O(x\log^2x)$ at zero and decays at infinity.
The second identity follows by rescaling $x$ in the integral. Solving
Eq.~\eqref{sm:a2} for its integral gives
\be
 J(2,z)=\frac{\pi^2A(z)}{z^2}
        -\frac{2\zeta(3)}{z^3}+\frac{\zeta(3)}8\,.
 \label{eq:J_from_A}
\ee

Now use $\log\cosh x=x-\log2+L_4(x)-L_2(x)$ and
$\int_0^\infty xL_k(x)\,\rd x=-\zeta(3)/k^2$ in
Eq.~\eqref{sm:CB}. This gives
\be
 \frac{C_{\mathrm B}(k)}{2\pi}
 =-\frac{k^2\zeta(3)}{32\pi^2}
  -\frac{\zeta(3)}{\pi^2k}
  +\frac{k}{\pi^2}[D(4,k)-D(2,k)]\,.
 \label{eq:CB_exchange_step}
\ee
Applying Eqs.~\eqref{eq:exchangelemma}--\eqref{eq:J_from_A}
produces the exact identity
\be
 \frac{C_{\mathrm B}(k)}{2\pi}
 =A(k)-A(k/2)+\frac{k}{2}A(4/k)
             -\frac{k}{4}A(8/k)-\frac{k^2\zeta(3)}{8\pi^2}\,,
 \label{eq:exchange}
\ee
which is Ref.~\cite[Eq.~(32)]{Hong:2026zul} in the present normalization.
No Bethe-root input is used in this integral manipulation.

The integer-level evaluations of $A$
\cite[Eq.~(3.14)]{Hatsuda:2014vsa} include
\be
 \begin{aligned}
 A(1)&=-\frac{\zeta(3)}{8\pi^2}+\frac{\log2}{4}\,,\qquad
 A(4)=-\frac{\zeta(3)}{4\pi^2}-\frac{\log2}{2}\,,\\
 A(8)&=-\frac{\zeta(3)}{8\pi^2}-\frac{7\log2}{4}\,.
 \end{aligned}
 \label{eq:Ainteger}
\ee
At $k=1$, Eq.~\eqref{eq:exchange} therefore reduces to
\be
 \frac{C_{\mathrm B}(1)}{2\pi}
 =-A(1/2)+\frac7{16}\log2-\frac{11\zeta(3)}{32\pi^2}\,.
 \label{eq:Ahalf_relation}
\ee
The analytic ABJM integral theorem gives
$C_{\mathrm B}(1)=-G-5\zeta(3)/(8\pi)$
\cite[Theorem~3.1 and Appendix~A.9]{Hosseini:2026lvj}.
Substituting this value and then Eq.~\eqref{eq:a2definition} yields
\be
 \boxed{
 A(1/2)=\frac G{2\pi}+\frac7{16}\log2-\frac{\zeta(3)}{32\pi^2}\,,
 \qquad
 a_2=-\frac G{4\pi}+\frac{7\zeta(3)}{64\pi^2}\,.}
 \label{eq:a2}
\ee
These are identities for the integral-defined functions. They remove
the need for an additional mirror-constant assumption in evaluating
Eq.~\eqref{eq:constant}. In particular,
\be
 \boxed{C(N_f)=\frac12C_{\mathrm B}(N_f)+\frac14C_{\mathrm B}(2N_f)
 +N_f^2\left(\frac{5\zeta(3)}{16\pi}-\frac G2\right)\,.}
 \label{eq:constantfinite}
\ee
The remaining physical assertion is that this analytic function is
the rank-independent term of the large-$N$ Bethe observable.

\subsection{Finite Clausen formulas at every integer flavor}\label{app:parity}
The ABJM finite formulas needed in
Eq.~\eqref{eq:constantfinite} are recalled with their exact
normalization. For odd positive $p$,
\bea
 C_{\mathrm B}(p)={}&
 -\left(\frac{p^2}{4}+\frac{3}{8p}\right)\frac{\zeta(3)}\pi
 +(-1)^{(p+1)/2}G
 \\
 &-\sum_{j=1}^{p-1}\left[
 \frac38(1-(-1)^j)j\Cltwo\!\left(\frac{2\pi j}{p}\right)
 +\frac14(-1)^j j\Cltwo\!\left(\frac{4\pi j}{p}\right)\right]\,,
 \label{eq:CBodd}
\eea
whereas, for every positive integer $p$,
\be
 C_{\mathrm B}(2p)=
 -\left[p^2-\frac{1+7(-1)^p}{4p}\right]\frac{\zeta(3)}\pi
 -\frac12\sum_{j=1}^{p-1}[3-(-1)^{p+j}]
          j\Cltwo\!\left(\frac{2\pi j}{p}\right)\,.
 \label{eq:CBeven}
\ee
These expressions were proposed in Ref.~\cite{Hosseini:2026dkj} and proved
as evaluations of Eq.~\eqref{sm:CB} in
Ref.~\cite[Theorem~3.1 and Appendix~A]{Hosseini:2026lvj}.
The proof uses root-of-unity evaluations of convergent Euler sums;
its validity does not depend on identifying the integral with a
finite-rank ABJM observable.

\begin{proposition}[The finite ADHM constant map]\label{prop:constant}
For the function $C(N_f)$ defined by Eq.~\eqref{eq:constant}, every
odd $N_f$ obeys
\bea
 C(N_f)={}&-\left(\frac{N_f^2}{16}+\frac9{16N_f}\right)
                         \frac{\zeta(3)}\pi
 +\frac{(-1)^{(N_f+1)/2}-N_f^2}{2}G
 \\
 &-\frac1{16}\sum_{j=1}^{N_f-1}(9-(-1)^j)
               j\Cltwo\!\left(\frac{2\pi j}{N_f}\right)
 -\frac18\sum_{j=1}^{N_f-1}(-1)^j
               j\Cltwo\!\left(\frac{4\pi j}{N_f}\right)\,,
 \label{sm:Codd}
\eea
and every even $N_f$ obeys
\bea
 C(N_f)={}&\left[-\frac{N_f^2}{16}
       +\frac{3+7(-1)^{N_f/2}}{4N_f}\right]\frac{\zeta(3)}\pi
       -\frac{N_f^2}{2}G
 \\
 &-\frac14\sum_{j=1}^{N_f/2-1}(5-(-1)^{N_f/2+j})
                  j\Cltwo\!\left(\frac{4\pi j}{N_f}\right)
 -\frac12\sum_{r=0}^{N_f/2-1}(2r+1)
                \Cltwo\!\left(\frac{2(2r+1)\pi}{N_f}\right)\,.
 \label{sm:Ceven}
\eea
Empty sums vanish.
\end{proposition}
\begin{proof}
For odd $N_f$, substitute Eq.~\eqref{eq:CBodd} with $p=N_f$ and
Eq.~\eqref{eq:CBeven} with $p=N_f$ into
Eq.~\eqref{eq:constantfinite}. The coefficients of
$j\Cltwo(2\pi j/N_f)$ combine as
\be
 -\frac3{16}(1-(-1)^j)-\frac18(3+(-1)^j)
 =-\frac1{16}(9-(-1)^j)\,.
\ee
The remaining Clausen sum has coefficient $-(-1)^j/8$.
The zeta and Catalan terms combine to those in Eq.~\eqref{sm:Codd}.

For even $N_f$, apply Eq.~\eqref{eq:CBeven} to both ABJM terms.
Split the sum from $C_{\mathrm B}(2N_f)$ into even indices $2j$
and odd indices $2r+1$. The even part contributes
$-j\Cltwo(4\pi j/N_f)/2$, which combines with the sum from
$C_{\mathrm B}(N_f)/2$ to give the coefficient
$-[5-(-1)^{N_f/2+j}]/4$.
The odd part has coefficient $-(2r+1)/2$.
Combining the remaining constants gives Eq.~\eqref{sm:Ceven}.
\end{proof}

Table~\ref{tab:specialconstants} illustrates the arithmetic structure of $C(N_f)$ through several low-flavor values.
The reductions use periodicity, oddness, and
$\Cltwo(2\theta)=2\Cltwo(\theta)+2\Cltwo(\theta+\pi)$,
obtained by retaining the even Fourier modes of Eq.~\eqref{eq:Clausen}.
In particular $C(1)=C_{\mathrm B}(1)$ follows analytically for the
defined constants, consistently with the mirror relation. Equality
of the full finite-rank observables requires the nonperturbative
terms as well.
\begin{table}[htp!]
\centering
\begin{tabular}{cl}
\toprule
$N_f$ & $C(N_f)$\\
\midrule
$1$ & $-G-\frac{5\zeta(3)}{8\pi}$\\[2pt]
$2$ & $-2G-\frac{3\zeta(3)}{4\pi}$\\[2pt]
$3$ & $-4G-\frac{3\zeta(3)}{4\pi}$\\[2pt]
$4$ & $-7G-\frac{3\zeta(3)}{8\pi}$\\[2pt]
$6$ & $5\Cltwo\!\big(\frac{2\pi}{3}\big)-18G
       -\frac{29\zeta(3)}{12\pi}$\\[2pt]
$7$ & $\frac32\Cltwo\!\big(\frac{2\pi}{7}\big)
       +3\Cltwo\!\big(\frac{4\pi}{7}\big)
       +\Cltwo\!\big(\frac{6\pi}{7}\big)
       -24G-\frac{22\zeta(3)}{7\pi}$\\[2pt]
$8$ & $4\Cltwo\!\big(\frac{\pi}{4}\big)
       -\frac{59G}{2}
       -\frac{59\zeta(3)}{16\pi}$\\
\bottomrule
\end{tabular}
\caption{Special values of the constant $C(N_f)$ for selected values of $N_f$.}
\label{tab:specialconstants}
\end{table}

Appendix~\ref{app:constants} supplies a separate finite
multiple-polylogarithm representation obtained directly from the
ADHM integral. It makes the roots of unity entering the constant
explicit before the Clausen reduction. The finite formulas, the
integral, and the large-flavor expansion derived below provide
independent forms suitable for numerical checks, with the analytic
status specified by Proposition~\ref{prop:constant}.

\section{Reconstruction of the exponential sectors}\label{sec:reconstruction}
The perturbative subtraction leaves a small but structured signal. At fixed $N_f$ the natural variable is $q=\e^{-t}$, with $t$ defined in Eq.~\eqref{eq:shift}. Resolving that signal requires both a precise critical value and the additive normalization of $C(N_f)$. An error in the constant does not decay with rank and eventually masks every exponential sector.

The physical remainder is defined without any continuation assumption by
\be
 \W_A^{\rm np}(N,N_f)=\W_{\rm A}(N,N_f)-\W_{\rm A}^{\rm pert}(N,N_f)\,,
 \label{eq:physical-remainder}
\ee
and is determined numerically by the Bethe solution. The decimal data allow a finite-sector model to be fitted, but do not give exact rational coefficients of its all-order expansion. The closed expression proposed below is therefore a conjecture for this remainder, even when its numerical discrepancy lies below the uncertainty in the critical values.

\subsection{Unconstrained finite-sector fits}

The reconstruction at $N_f=2,4$ fits
\be
 \frac{2\pi}{N_f^2}[\W_{\rm A}-\W_{\rm A}^{\rm pert}]
 \simeq\sum_{m=1}^{K}(A_mt+B_m)q^m\,,\qquad K=16\,.\label{sm:fit}
\ee
All $32$ parameters are free. We solve the linear least-squares problem by QR factorization~\cite[Section~5.3]{GolubVanLoan:2013}.\footnote{QR factorization solves an overdetermined linear fit. PSLQ instead searches for integer relations among supplied real numbers~\cite{ferguson1999analysis}; it is not used in these sector fits.} The fits use $46$ even training ranks and $45$ odd validation ranks in $10$--$100$, or $41$ even training ranks and $40$ odd validation ranks in $20$--$100$, at $950$-digit working precision. The windows overlap and are not independent data sets. With $q_0=q(N_{\min})$, the target is divided by $q$ and the columns are rescaled to $t(q/q_0)^{m-1}$ and $(q/q_0)^{m-1}$. The ansatz assumes integer multiples of the prescribed action and degree-one polynomials in that action. Neither the relation $B_m=A_m/m$ nor a divisor law is imposed during the fit. Both are compared only after the coefficients have been obtained.

In the later window, the differences $A_{16}-c_{16}^{(N_f)}$ are approximately $1.89\times10^{-10}$ and $-4.09\times10^{-7}$ at flavors two and four. The different diagnostic $16B_{16}/A_{16}-1$ is approximately $6.45\times10^{-9}$ and $-7.69\times10^{-5}$. Neither diagnostic measures $950$-digit physical coefficient accuracy.

The maximum relative residual on the withheld odd ranks is $3.84\times10^{-204}$ at $N_f=2$ and $7.83\times10^{-147}$ at $N_f=4$ in the $20$--$100$ window. These residuals measure the finite model's agreement with the withheld critical values. The fitted coefficients also contain finite-window and omitted-sector effects, so their accuracy is not set by the working precision alone.

The first sector omitted from a $K=16$ fit has $m=17$. Under the divisor law below its normalized contribution is $c_{17}^{(N_f)}(t+1/17)q^{17}$, with
\be
 c_{17}^{(2)}=-\frac{4914}{289}\,,\qquad
 c_{17}^{(4)}=\frac{2457}{289}\,.
 \label{eq:first-omitted}
\ee
These are coefficients of the proposed continuation, not independent extractions. A finite fit allows omitted sectors to shift its retained coefficients, most strongly near the cutoff. The two fitting windows provide a measure of this sensitivity without changing the arithmetic precision or imposing the proposed coefficient law.

\subsection{Rational reconstruction of the fitted coefficients}
We apply continued-fraction rational approximation separately to the fitted $A_m$ and $B_m$, restricting the denominator to at most $10^6$. A value is retained only when both fitting windows select the same rational and this is the unique rational in an empirical interval determined by their spread and the $200$-significant-digit export resolution. The interval and uniqueness test are specified in Appendix~\ref{app:numerics}. Neither the divisor law nor the relation between $A_m$ and $B_m$ enters this step.

Table~\ref{tab:recovered-coefficients} gives the resulting rational values through $m=15$ at both flavors. Both underlying fits retain $K=16$ sectors; the four coefficients at $m=16$ fail the recovery criterion and are not tabulated. Every recovered pair satisfies $B_m=A_m/m$, and comparison with the proposed divisor laws below gives $A_m=c_m^{(N_f)}$. These equalities are exact statements about the reconstructed rationals. Their identification with the physical asymptotic coefficients remains conditional on rationality in the specified class and on the empirical interval containing the physical coefficient. The overlapping windows do not supply a rigorous estimate of a shared truncation bias.

\begin{table}[htp!]
\centering
\renewcommand{\arraystretch}{1.35}
\begin{tabular}{ccccc}
\toprule
 & \multicolumn{2}{c}{$N_f=2$} & \multicolumn{2}{c}{$N_f=4$} \\
\cmidrule(lr){2-3}\cmidrule(lr){4-5}
$m$ & $A_m$ & $B_m$ & $A_m$ & $B_m$ \\
\midrule
1 & $-1$ & $-1$ & $\frac{1}{2}$ & $\frac{1}{2}$ \\
2 & $-\frac{9}{4}$ & $-\frac{9}{8}$ & $\frac{1}{4}$ & $\frac{1}{8}$ \\
3 & $-\frac{28}{9}$ & $-\frac{28}{27}$ & $\frac{14}{9}$ & $\frac{14}{27}$ \\
4 & $-\frac{57}{16}$ & $-\frac{57}{64}$ & $-\frac{9}{16}$ & $-\frac{9}{64}$ \\
5 & $-\frac{126}{25}$ & $-\frac{126}{125}$ & $\frac{63}{25}$ & $\frac{63}{125}$ \\
6 & $-7$ & $-\frac{7}{6}$ & $\frac{7}{9}$ & $\frac{7}{54}$ \\
7 & $-\frac{344}{49}$ & $-\frac{344}{343}$ & $\frac{172}{49}$ & $\frac{172}{343}$ \\
8 & $-\frac{441}{64}$ & $-\frac{441}{512}$ & $-\frac{57}{64}$ & $-\frac{57}{512}$ \\
9 & $-\frac{757}{81}$ & $-\frac{757}{729}$ & $\frac{757}{162}$ & $\frac{757}{1458}$ \\
10 & $-\frac{567}{50}$ & $-\frac{567}{500}$ & $\frac{63}{50}$ & $\frac{63}{500}$ \\
11 & $-\frac{1332}{121}$ & $-\frac{1332}{1331}$ & $\frac{666}{121}$ & $\frac{666}{1331}$ \\
12 & $-\frac{133}{12}$ & $-\frac{133}{144}$ & $-\frac{7}{4}$ & $-\frac{7}{48}$ \\
13 & $-\frac{2198}{169}$ & $-\frac{2198}{2197}$ & $\frac{1099}{169}$ & $\frac{1099}{2197}$ \\
14 & $-\frac{774}{49}$ & $-\frac{387}{343}$ & $\frac{86}{49}$ & $\frac{43}{343}$ \\
15 & $-\frac{392}{25}$ & $-\frac{392}{375}$ & $\frac{196}{25}$ & $\frac{196}{375}$ \\
\bottomrule
\end{tabular}
\caption{Rational coefficients recovered separately from the freely fitted $A_m$ and $B_m$ at $N_f=2,4$. The two $K=16$ fits use even ranks in $10\leq N\leq100$ and $20\leq N\leq100$, with odd ranks withheld. Each numerical entry is the unique rational with denominator at most $10^6$ in its empirical window-stability interval, and the two fits select the same rational. The interval is defined in the text and is not a rigorous error bound on the physical coefficient. Neither $B_m=A_m/m$ nor the divisor formula is imposed in the recovery; both relations hold for every displayed pair after recovery. The fitted $m=16$ coefficients fail the recovery criterion and are omitted from the table.}
\label{tab:recovered-coefficients}
\end{table}

\subsection{The proposed divisor laws}

Let $\sigma_3(m)=\sum_{d\mid m}d^3$ and set $\sigma_3(x)=0$ when $x$ is not a positive integer. The three proposed sequences are
\bea
 m^2c_m^{(1)}&=5(-1)^{m+1}\big[\sigma_3(m)-4\sigma_3(m/2)\big]\,,\\
 m^2c_m^{(2)}&=-\sigma_3(m)+16\sigma_3(m/4)\,,\\
 m^2c_m^{(4)}&=\frac12\big[\sigma_3(m)-7\sigma_3(m/2)-28\sigma_3(m/4)+64\sigma_3(m/8)\big]\,.\label{eq:divisor}
\eea
Flavor one uses the ABJM level-one completion~\cite{Hosseini:2026dkj,Hosseini:2026lvj}; the flavor-two and flavor-four expressions are reconstructed from the finite fits described above. The first proposed coefficients are $5,-1,1/2$, respectively, so the first correction at flavor two has the opposite sign to the other two.

\begin{conjecture}[Special-flavor completion]\label{conj:completion}
For $N_f=1,2,4$ and positive integer $N$, the critical value on the branch of Section~\ref{sec:branch} satisfies
\be
 \boxed{\begin{aligned}
 \W_{\rm A}(N,N_f)&=\W_{\rm A}^{\rm pert}(N,N_f)+\W_A^{\rm np}(N,N_f)\,,\\
 \W_A^{\rm np}(N,N_f)&=\frac{N_f^2}{2\pi}\sum_{m\geq1}c_m^{(N_f)}
 \left(t+\frac1m\right)q^m\,.
 \end{aligned}}\label{eq:completion}
\ee
The constant is given by Eq.~\eqref{eq:constantfinite}, the action by Eq.~\eqref{eq:shift}, and the coefficients by Eq.~\eqref{eq:divisor}.
\end{conjecture}
The flavor-one proposal is inherited from the level-one ABJM completion after matching the branch normalization. For flavors two and four, the finite fits support both the divisor combinations and the relative coefficient $B_m=A_m/m$. Extending them to all sectors is the conjectural step. In particular, neither a successful QR fit nor a small held-out error is an integer-relation certificate. Analytic identities derived from Eq.~\eqref{eq:completion} below hold exactly for its explicit series. Their interpretation as identities of the Bethe observable is conditional on this conjecture.

At $N_f=2$ the alternating ABJM sign pattern is absent. At $N_f=4$ contributions supported on multiples of $2,4,8$ modify the sequence on even sectors. This dyadic structure leads to the modular levels derived next.

Once the constant and divisor law are fixed, no parameter is adjusted in the $897$ special-flavor comparisons. Low ranks outside the fit windows require substantially more than sixteen sectors before the sum reaches the precision of the critical values. These comparisons test the all-order expression beyond the coefficient range used to reconstruct it. The numerical analysis in Section~\ref{sec:validation} specifies their accuracy and the number of sectors they resolve.

\section{Eisenstein primitives, modular levels and convergence}\label{sec:modular}
\subsection{From divisor coefficients to Lambert series}
The divisor laws in Eq.~\eqref{eq:divisor} admit an arithmetic description that is independent of their proposed identification with the Bethe observable. Throughout this section $N_f\in\{1,2,4\}$. We first regard $q$ as a complex variable in the open unit disk, and impose the physical relation $q=\e^{-t}$ only when evaluating the correction at real positive action.

The nome $q$ is constructed from the rank and flavor number after solving the Bethe problem. It is not a microscopic fugacity in Eq.~\eqref{eq:exponentiated}, whose rational dependence on the holonomies exhibits no modular transformation in this variable. The modular description therefore supplies information beyond the form of the defining equations. It organizes the resolved exponential coefficients into Eisenstein series with specific levels and fixes an infinite set of relations among the proposed sectors.

Define the Lambert series and the normalized weight-four Eisenstein series by
\be
 \cL_3(z)=\sum_{r\geq1}\Li_3(z^r)
 =\sum_{m\geq1}\frac{\sigma_3(m)}{m^3}z^m\,,
 \qquad
 E_4(z)=1+240\sum_{m\geq1}\sigma_3(m)z^m\,.
 \label{eq:lambert}
\ee
Arguments of $E_4$ are nomes in this notation. To verify the Lambert rearrangement, expand $\Li_3(z^r)=\sum_{s\geq1}z^{rs}/s^3$. For $|z|<1$ the double series converges absolutely, since
$\sum_{r,s\geq1}|z|^{rs}/s^3\leq |z|/(1-|z|)^2$.
The coefficient of $z^m$ is therefore
$\sum_{s\mid m}s^{-3}=\sigma_3(m)/m^3$.
For a positive integer $d$, the coefficient of $q^m$ in $\cL_3(q^d)$ is $d^3\sigma_3(m/d)/m^3$, with the established convention that the divisor sum vanishes when $d\nmid m$.

The three primitives are
\bea
 \Phi_1(q)&=-\frac5{2\pi}\left[\cL_3(-q)-\frac12\cL_3(q^2)\right]\,,\\
 \Phi_2(q)&=-\frac2\pi\left[\cL_3(q)-\frac14\cL_3(q^4)\right]\,,\\
 \Phi_4(q)&=\frac4\pi\left[\cL_3(q)-\frac78\cL_3(q^2)-\frac7{16}\cL_3(q^4)+\frac18\cL_3(q^8)\right]\,.
 \label{eq:modular}
\eea
The coefficient rule above gives
$[q^m]\Phi_{N_f}=N_f^2c_m^{(N_f)}/(2\pi m)$.%
\footnote{The brackets denote a Taylor coefficient.}
With $D=q\partial_q$, this proves
\be
 \W_A^{\rm np}(N,N_f)
 =\left.(1+tD)\Phi_{N_f}(q)\right|_{q=\e^{-t}}\,.
 \label{eq:primitive_tail}
\ee
In particular, the derivative acts on $\Phi_{N_f}$ while $t$ is held fixed. On the physical ray the same operation is $1-t\partial_t$ applied to $\Phi_{N_f}(\e^{-t})$. Distinguishing these two uses of the action is necessary when comparing tails at rescaled or analytically continued arguments.

\subsection{The Fourier primitive and its constant mode}
For a Fourier series $h(q)=a_0+\sum_{m\geq1}a_mq^m$ convergent for $|q|<1$, define
$D^{-3}(h-a_0)=\sum_{m\geq1}a_mq^m/m^3$.
This prescription fixes zero constant term and excludes polynomials in $\log q$. It selects the primitive that is holomorphic at $q=0$ and vanishes there. If $q=\e^{2\pi\ii\tau}$ with $\tau$ in the upper half-plane, then
$D=(2\pi\ii)^{-1}\partial_\tau$, and the same primitive can be written as
\be
 D^{-3}(h-a_0)
 =-\frac{(2\pi\ii)^3}{2}
 \int_{\tau}^{\ii\infty}(z-\tau)^2
 \bigl[h(\e^{2\pi\ii z})-a_0\bigr]\,\rd z\,.
 \label{eq:eichler_integral}
\ee
The path is vertical. Exponential decay after removal of $a_0$ permits termwise integration. Each Fourier mode contributes
$-2\e^{2\pi\ii m\tau}/(2\pi\ii m)^3$, which gives the stated coefficient after multiplication by the prefactor. Equivalently, three derivatives of the integral give $-2(h-a_0)$.

For a dilation, the chain rule gives
\be
 D^3\cL_3(q^d)=\frac{d^3}{240}[E_4(q^d)-1]\,.
 \label{sm:dilation}
\ee
The ABJM primitive convention consequently reads~\cite[Eq.~(14)]{Hosseini:2026dkj}
\bea
 \widetilde E_L(q)
 &=D^{-3}\bigl[E_4(q)-L^2E_4(q^L)-(1-L^2)\bigr]\\
 &=240\bigl[\cL_3(q)-L^{-1}\cL_3(q^L)\bigr]\,.
 \label{sm:eichler}
\eea
Here $L$ is a positive integer dilation. The subtraction in Eq.~\eqref{sm:eichler} is essential: a nonzero constant has no holomorphic $D^{-3}$ primitive at the origin. This is the zero-mode convention appropriate to a noncuspidal Eichler integral~\cite[Eq.~(8.1)]{Pasol:2013}. It does not make the primitive, or the expression obtained after applying $1+tD$, a modular form.

\subsection{Physical and shifted modular coordinates}
The alternating coefficients at flavor one are naturally expressed in
\be
 q=\e^{2\pi\ii\tau}\,,\qquad
 Q=-q=\e^{2\pi\ii\tau_2}\,,\qquad
 \tau_2=\tau+\frac12\,.
\ee
The logarithmic derivative is unchanged,
$D=q\partial_q=Q\partial_Q$.
The half-period identity is~\cite[Eq.~(2.56)]{Dorigoni:2020oon}
\be
 E_4(-q)=-E_4(q)+18E_4(q^2)-16E_4(q^4)\,.
 \label{sm:halfperiod}
\ee
That reference uses $G_4=2\zeta(4)E_4$; division by its constant mode gives Eq.~\eqref{sm:halfperiod} in our normalization. A direct coefficient proof follows from
$\sigma_3(2r)=9\sigma_3(r)-8\sigma_3(r/2)$.
For $r=2^a s$ with $s$ odd, this identity follows by substituting
$\sigma_3(r)=(8^{a+1}-1)\sigma_3(s)/7$.
The even Fourier coefficients of Eq.~\eqref{sm:halfperiod} then agree, and its odd coefficients and constant term agree separately.

Applying Eq.~\eqref{sm:dilation} to the primitives gives
\bea
 96\pi D^3\Phi_1
 &=E_4(q)-14E_4(q^2)+16E_4(q^4)-3\,,\\
 -120\pi D^3\Phi_2
 &=E_4(q)-16E_4(q^4)+15\,,\\
 60\pi D^3\Phi_4
 &=E_4(q)-7E_4(q^2)-28E_4(q^4)+64E_4(q^8)-30\,.
 \label{sm:thirdderivatives}
\eea
Let $h_{N_f}(\tau)$ denote the full Eisenstein combinations on these right-hand sides, before the constant subtraction. Their constant modes are $3,-15,30$. In the shifted coordinate define
$h_{N_f}^{\rm sh}(\tau_2)=h_{N_f}(\tau_2-1/2)$.
Equation~\eqref{sm:halfperiod} yields
\bea
 h_1^{\rm sh}(\tau_2)&=-E_4(Q)+4E_4(Q^2)\,,\\
 h_2^{\rm sh}(\tau_2)&=-E_4(Q)+18E_4(Q^2)-32E_4(Q^4)\,,\\
 h_4^{\rm sh}(\tau_2)&=-E_4(Q)+11E_4(Q^2)-44E_4(Q^4)+64E_4(Q^8)\,.
 \label{eq:shifted_full_forms}
\eea
The same constant modes occur in both coordinates.

For a positive integer $L$, let $\Gamma_0(L)$ be the subgroup of
$\mathrm{SL}_2(\mathbb Z)$ whose lower-left entry is divisible by $L$.
The weight-four slash operation is
\be
 (h|_4\gamma)(\tau)=(\mathsf c\tau+\mathsf d)^{-4}h(\gamma\tau)\,,
 \qquad
 \gamma=\begin{pmatrix}\mathsf a&\mathsf b\\\mathsf c&\mathsf d\end{pmatrix}\,.
\ee
The space $M_4(\Gamma_0(L))$ consists of functions holomorphic on the upper half-plane and at every cusp, invariant under this weight-four slash operation for all $\gamma\in\Gamma_0(L)$. Each $E_4(\e^{2\pi\ii d\tau})$ is a holomorphic weight-four form for $\Gamma_0(d)$. Indeed, for
$\gamma=\left(\begin{smallmatrix}\mathsf a&\mathsf b\\\mathsf c&\mathsf d\end{smallmatrix}\right)\in\Gamma_0(d)$,
the matrix
$\left(\begin{smallmatrix}\mathsf a&d\mathsf b\\\mathsf c/d&\mathsf d\end{smallmatrix}\right)$
is integral with determinant one. The level-one transformation law of $E_4$ then gives the required weight-four law, and the degeneracy construction supplies holomorphy at every cusp~\cite[Theorem~5.8]{Stein:2007}.
The convention in Ref.~\cite{Stein:2007} normalizes the coefficient of $q$ to one, so its weight-four Eisenstein series is $E_4/240$ here. This overall factor does not affect the level statement. It follows that the physical forms belong to levels $4,4,8$, while the shifted forms belong to levels $2,4,8$.

The coordinate dependence can also be expressed as a group conjugation. Put
$T_{1/2}=\left(\begin{smallmatrix}1&1/2\\0&1\end{smallmatrix}\right)$.
If a form is modular for $\Gamma_0(L)$ in $\tau_2=T_{1/2}\tau$, its pullback is modular for $T_{1/2}^{-1}\Gamma_0(L)T_{1/2}$ in $\tau$. For an integral matrix $\gamma$, one has
\be
 T_{1/2}\gamma T_{1/2}^{-1}
 =\begin{pmatrix}
 \mathsf a+\mathsf c/2&\mathsf b+(\mathsf d-\mathsf a)/2-\mathsf c/4\\\mathsf c&\mathsf d-\mathsf c/2
 \end{pmatrix}\,.
\ee
If this lies in $\Gamma_0(2)$, then $\mathsf c$ is even and $\mathsf a\mathsf d-\mathsf b\mathsf c=1$ makes $\mathsf a,\mathsf d$ odd. Integrality of the upper-right entry requires $4\mid\mathsf c$; this condition is also sufficient. Hence
\be
 \begin{split}
 \bigl(T_{1/2}^{-1}\Gamma_0(2)T_{1/2}\bigr)
 \cap\mathrm{SL}_2(\mathbb Z)&=\Gamma_0(4)\,,\\
 T_{1/2}^{-1}\Gamma_0(L)T_{1/2}&=\Gamma_0(L)
 \qquad(4\mid L)\,.
 \end{split}
 \label{eq:conjugated_group}
\ee
The second equality follows by applying the same integrality test to $T_{1/2}$ and its inverse. Thus the flavor-one level-two description uses the shifted coordinate. Its conjugate group in the physical coordinate also contains rational nonintegral matrices, and is not $\Gamma_0(2)$ with the coordinate left unchanged.

\subsection{Minimal levels and cusp constants}
Membership at a stated level does not establish minimality. To exclude every smaller integral level, consider
$h(\tau)=\sum_{d\mid8}a_dE_4(\e^{2\pi\ii d\tau})$ and
$\gamma_M=\left(\begin{smallmatrix}1&0\\M&1\end{smallmatrix}\right)$ for a positive integer $M$. Its cusp constant is
\be
 \lim_{\im\tau\to\infty}(h|_4\gamma_M)(\tau)
 =\sum_{d\mid8}a_d\frac{\gcd(d,M)^4}{d^4}\,.
 \label{sm:cusp}
\ee
For one summand, write $g=\gcd(d,M)$, $d'=d/g$, $M'=M/g$, and choose integers $\mathsf b,\mathsf d$ with $d'\mathsf d-M'\mathsf b=1$. The integral matrix with rows $(d',\mathsf b)$ and $(M',\mathsf d)$ maps
$w=(g\tau-\mathsf b)/d'$ to $d\gamma_M\tau$.
Its transformation law reduces the slashed summand to
$d'^{-4}E_4(\e^{2\pi\ii w})$.
Since $\im w\to\infty$, its limit is $(g/d)^4$, proving Eq.~\eqref{sm:cusp}.

If $h$ were modular for $\Gamma_0(M)$, then $\gamma_M$ would preserve it, so this cusp constant would equal its constant at infinity. Table~\ref{tab:cusp_levels} evaluates the necessary condition for all six forms. Only the exponent $v_2(M)$ of two in $M$ matters, because every dilation divides eight.
\begin{table}[htp!]
 \centering\small
 \begin{tabular}{ccrrrrrr}
 \toprule
 $N_f$ & Coordinate & $a_0$ & $v_2(M)=0$ & $v_2(M)=1$ & $v_2(M)=2$ & $v_2(M)\geq3$ & $L_{\min}$\\
 \midrule
 1 & $\tau$   & $3$   & $3/16$   & $-12$   & $3$   & $3$   & $4$\\
 2 & $\tau$   & $-15$ & $15/16$  & $0$     & $-15$ & $-15$ & $4$\\
 4 & $\tau$   & $30$  & $15/32$  & $-15/2$ & $-30$ & $30$  & $8$\\
 \addlinespace
 1 & $\tau_2$ & $3$   & $-3/4$   & $3$     & $3$   & $3$   & $2$\\
 2 & $\tau_2$ & $-15$ & $0$      & $15$    & $-15$ & $-15$ & $4$\\
 4 & $\tau_2$ & $30$  & $-15/32$ & $15/2$  & $-30$ & $30$  & $8$\\
 \bottomrule
 \end{tabular}
 \caption{Exact cusp constants from Eq.~\eqref{sm:cusp} for the full forms $h_{N_f}$ and $h_{N_f}^{\rm sh}$. The column $a_0$ is the constant at infinity. Equality with $a_0$ is necessary for membership in $M_4(\Gamma_0(M))$. Combined with the established memberships, the table proves the minimal integral levels in the last column.}
 \label{tab:cusp_levels}
\end{table}
For example, the physical flavor-four form requires $v_2(M)\geq3$, and membership at level eight has already been proved. The other five rows give the remaining minimality claims in the same manner. This argument covers arbitrary positive integral $M$, including levels that are not powers of two. It does not classify the full stabilizer in $\mathrm{SL}_2(\mathbb R)$.

All these statements concern the full forms. A nonzero constant is not a weight-four modular form, so subtracting $a_0$ generally destroys the homogeneous transformation law. The objects $D^3\Phi_{N_f}$ in Eq.~\eqref{sm:thirdderivatives} are constant-subtracted forms. Their convergent primitives are well defined by Eq.~\eqref{eq:eichler_integral}, but a transformation formula for a noncuspidal primitive must retain the contribution of its constant mode. No modular-invariance statement for the physical twisted superpotential follows from level membership alone.

\subsection{Convergence and explicit truncation control}
The elementary divisor bound
$\sigma_3(m)=m^3\sum_{d\mid m}d^{-3}\leq\zeta(3)m^3$
implies $|c_m^{(N_f)}|=O(m)$. Thus both series
$\sum_{m\geq1}c_m^{(N_f)}q^m$ and
$\sum_{m\geq1}c_m^{(N_f)}q^m/m$
converge absolutely in the open unit disk. The radius cannot exceed one. At each odd prime $\mathfrak p$, all proper dilation terms in the divisor laws vanish and
\be
 (c_\mathfrak p^{(1)},c_\mathfrak p^{(2)},c_\mathfrak p^{(4)})
 =(5,-1,\tfrac12)(\mathfrak p+\mathfrak p^{-2})\,.
 \label{sm:prime}
\ee
The $\mathfrak p$th roots of the magnitudes tend to one, both before and after division by $\mathfrak p$. Together with the polynomial upper bound, the Cauchy--Hadamard formula gives radius exactly one for each series.

The same subsequence shows that their terms fail to approach zero at every point of $|q|=1$. This proves divergence there as ordinary series, but does not prove a natural boundary or exclude analytic continuation through boundary arcs. The physical correction is not itself a Taylor series at zero: after $t=-\log q$ it reads $(1-\log q\,D)\Phi_{N_f}(q)$ on a specified logarithmic branch of the punctured disk.

For practical truncation, the divisor formulas give
$|c_m^{(N_f)}|\leq b_{N_f}\zeta(3)m$, where
\be
 (b_1,b_2,b_4)=\left(5,\frac54,\frac{39}{32}\right)\,,
 \qquad K_{N_f}=\frac{N_f^2\zeta(3)b_{N_f}}{2\pi}\,.
\ee
At flavor one, write $m=2^as$ with $s$ odd. The difference
$\sigma_3(m)-4\sigma_3(m/2)$ equals $\sigma_3(s)$ for $a=0$ and $(4\cdot8^a+3)\sigma_3(s)/7$ for $a\geq1$. It is therefore nonnegative and bounded above by $\sigma_3(m)$, giving $b_1=5$. The triangle inequalities at flavors two and four give
$1+16/4^3=5/4$ and
$(1+7/2^3+28/4^3+64/8^3)/2=39/32$.
For real $t>0$ and $q=\e^{-t}$, summing the omitted terms after sector $M$ therefore yields
\bea
 |\mathrm{tail}_{>M}|
 &\leq K_{N_f}\sum_{m=M+1}^{\infty}(tm+1)q^m\\
 &\leq K_{N_f}q^{M+1}
 \left[t\left(\frac{M+1}{1-q}+\frac{q}{(1-q)^2}\right)
 +\frac1{1-q}\right]\,.
 \label{sm:tailbound}
\eea
The two geometric sums in the second line are exact. This bound controls the truncation of the divisor series at every positive action.

\subsection{What finite data determine}
The modular description narrows the reconstruction to a particular function class. Within the four-dimensional Eisenstein span
\be
 \operatorname{span}_{\mathbb C}
 \{E_4(q),E_4(q^2),E_4(q^4),E_4(q^8)\}\,,
\ee
four appropriate nonconstant Fourier coefficients determine a form uniquely. For a difference of two forms, the coefficients of $q,q^2,q^4,q^8$ successively force the four dilation coefficients to vanish. With the primitive convention fixed, this also fixes its zero-constant primitive. The conclusion is conditional on membership in this specified span. It is neither uniqueness in the full space $M_4(\Gamma_0(8))$ nor a proof that the physical observable belongs to the Eisenstein ansatz.

A finite-data construction makes the distinction explicit. At one fixed flavor, let $q_a$ be the finite set of tested nomes and define, for arbitrary $\varepsilon$,
\be
 \Phi_\varepsilon(q)=\Phi(q)
 +\varepsilon q^{17}\prod_a(q-q_a)^2\,.
 \label{eq:countercontinuation}
\ee
The added polynomial preserves the first sixteen coefficients. Its value and logarithmic derivative vanish at every $q_a$, so all tested values of $(1+tD)\Phi$ are also preserved exactly. A polynomial addition leaves the radius of convergence unchanged. These alternatives are not asserted to be physical or modular; they show why finite coefficients and finite-rank agreement cannot establish either property without an additional argument. The analytic results of this section characterize the proposed continuation. Its identification with the selected Bethe critical value remains the conjecture tested numerically in Section~\ref{sec:validation}.

\section{The type-IIA expansion of the perturbative parent}\label{sec:genus}
\subsection{Fixed-coupling variables and genus normalization}
The large-flavor limit reorganizes the shifted-rank expression into a genus expansion. We retain the ABJM notation for the coupling and genus coefficients, with the ADHM shift fixed by its own rank:
\be
 \lambda=\frac{N}{N_f}\,,\qquad
 \lamh=\lambda+\frac7{24}\,,\qquad
 \Nh=N_f\lamh+\frac1{3N_f}\,.
 \label{eq:IIAvariables}
\ee
The limit is $N,N_f\to\infty$ with $\lambda>0$ fixed. In ABJM the corresponding shifted coupling is $\lambda-1/24$. The parent becomes
\be
 \W_{\rm A}^{\rm pert}(N,N_f)
 =\frac{\pi^2\sqrt2}{6}N_f^2
       \left(\lamh+\frac1{3N_f^2}\right)^{3/2}+C(N_f)\,.
 \label{sm:genusparent}
\ee
Following Refs.~\cite[Eqs.~(7)--(9)]{Hosseini:2026dkj} and \cite[Section~3.3]{Hosseini:2026lvj}, we define its genus coefficients by
\be
 \W_{\rm A}^{\rm pert}(N,N_f)
 \sim-\sum_{h\geq0}(2\pi\ii\lambda)^{2h-2}\cW_h(\lambda)N^{2-2h}\,.
 \label{eq:genusconvention}
\ee
Because $N=N_f\lambda$, the summand including its leading minus sign is
$(-1)^h(2\pi)^{2h-2}\cW_h(\lambda)N_f^{2-2h}$.
This identity fixes the signs and powers of $2\pi$ in the following formulas. The $\cW_h$ are real coefficients of this parent expansion.

\subsection{Large-flavor expansion of the constant}
The constant integral~\eqref{eq:constant} gives the coupling-independent contribution at each genus. Define Bernoulli numbers by
$z/(\e^z-1)=\sum_{r\geq0}\mathrm B_rz^r/r!$.
Expanding the kernel at the endpoint $x=0$ gives
\bea
 \log\cosh x
 &=\sum_{j\geq1}\frac{2^{2j}(2^{2j}-1)\mathrm B_{2j}}
                      {2j(2j)!}x^{2j}\,,\\
 [x^{2j}]\left[\frac12\log\cosh x+\frac14\log\cosh\frac x2\right]
 &=\frac{(4^j-1)(2\cdot4^j+1)\mathrm B_{2j}}{8j(2j)!}\,.
 \label{sm:kernelcoefficient}
\eea
Expanding $\log(1-\e^{-N_fx})$ into exponentials and integrating each monomial yields
\be
 \int_0^\infty x^{2j}\log(1-\e^{-N_fx})\,\rd x
 =-\frac{(2j)!\zeta(2j+2)}{N_f^{2j+1}}\,.
 \label{eq:log-moment}
\ee
Absolute convergence of the positive absolute-value series justifies this interchange. Substitution of the even-zeta identity
$\zeta(2r)=(-1)^{r+1}\mathrm B_{2r}(2\pi)^{2r}/[2(2r)!]$
then gives
\be
 \frac{C(N_f)}{2\pi}\sim a_2N_f^2-\frac{\log2}{8}
 +\sum_{j\geq1}\left(\frac{2\pi}{N_f}\right)^{2j}
 \frac{(-1)^j(4^j-1)(2\cdot4^j+1)|\mathrm B_{2j}\mathrm B_{2j+2}|}
 {4j(2j+2)!}\,.
 \label{sm:constantseries}
\ee
This is the normalization of the kernel expansion in Ref.~\cite[Appendix~A]{Hong:2026zul}. In particular, the first inverse-flavor terms in $C(N_f)/(2\pi)$ are
\be
 -\frac{\pi^2}{160N_f^2}
 +\frac{11\pi^4}{10080N_f^4}
 -\frac{43\pi^6}{50400N_f^6}\,.
 \label{eq:first-constant-terms}
\ee

The expansion holds in the Poincar\'e asymptotic sense at each fixed order. To prove this, truncate the kernel after $x^{2J}$ and split the integral at $0<\delta<\pi/2$, inside the nearest singularity of $\log\cosh x$. On $[0,\delta]$, its analytic Taylor remainder is uniformly $O(x^{2J+2})$, so Eq.~\eqref{eq:log-moment} bounds its contribution by $O(N_f^{-2J-2})$. On $[\delta,\infty)$,
\[
 |\log(1-\e^{-N_fx})|\leq
 \frac{\e^{-N_fx}}{1-\e^{-N_f\delta}}\,,
\]
and the exact kernel grows at most linearly. This part and the tails of the finitely many Taylor monomials are exponentially small. The argument proves an asymptotic expansion of the convergent integral.

\subsection{Coefficients at arbitrary genus}
The coefficient of $N_f^{2-2h}$ from the shifted power in Eq.~\eqref{sm:genusparent} is
\be
 \frac{\pi^2\sqrt2}{6}\binom{3/2}{h}
       3^{-h}\lamh^{3/2-h}\,.
 \label{eq:binomial-genus}
\ee
The constant contributes $2\pi a_2N_f^2$, $-\pi\log2/4$, and the inverse powers in Eq.~\eqref{sm:constantseries}. If $b_h$ denotes the sum of the two contributions at order $N_f^{2-2h}$, Eq.~\eqref{eq:genusconvention} gives
$\cW_h=(-1)^h(2\pi)^{2-2h}b_h$.
Thus
\bea
 \cW_0(\lambda)
 &=\frac{2\sqrt2\pi^4}{3}\lamh^{3/2}
       -2\pi^2G+\frac{7\pi\zeta(3)}8\,,\\
 \cW_1(\lambda)
 &=-\frac{\pi^2}{6\sqrt2}\lamh^{1/2}+\frac\pi4\log2\,,
 \label{eq:genus01}
\eea
where the analytic evaluation of $a_2$ in Eq.~\eqref{eq:a2} has been used. For $h=j+1\geq2$,
\bea
 \cW_{j+1}(\lambda)
 &=\frac{(-1)^{j+1}\sqrt2}{2^{2j+1}3^{j+2}\pi^{2j-2}}
       \binom{3/2}{j+1}\lamh^{1/2-j}+\omega_j\,,\\
 \omega_j
 &=-\frac{\pi(4^j-1)(2\cdot4^j+1)
                |\mathrm B_{2j}\mathrm B_{2j+2}|}{2j(2j+2)!}\,.
 \label{eq:genus}
\eea
For example,
\bea
 \cW_2&=\frac{\sqrt2}{576}\lamh^{-1/2}-\frac{\pi}{320}\,,\\
 \cW_3&=\frac{\sqrt2}{41472\pi^2}\lamh^{-3/2}
                       -\frac{11\pi}{80640}\,,\\
 \cW_4&=\frac{\sqrt2}{1327104\pi^4}\lamh^{-5/2}
                       -\frac{43\pi}{1612800}\,.
 \label{eq:genus-examples}
\eea

At equal shifted coupling, the binomial contribution at genus $h$ is $2^{-h}$ times the ABJM one. For the coupling-independent pieces in the convention of Ref.~\cite[Eq.~(9)]{Hosseini:2026dkj},
\be
 \frac{\omega_j^{\rm A}}{\omega_j^{\rm B}}
 =\frac12+4^{-j-1}\,.
 \label{eq:genus-ratio}
\ee
The theory superscripts in this comparison distinguish the two coefficient families. This relation follows by division of their Bernoulli expressions and concerns the parents, not their full finite-coupling observables.

\subsection{Divergence and the order of limits}
Using
$|\mathrm B_{2r}|=2(2r)!\zeta(2r)/(2\pi)^{2r}$
in Eq.~\eqref{eq:genus} gives the exact expression
\be
 \omega_j=-\frac{(2j)!}{j\pi^{4j+1}}
  (1-4^{-j})\left(1+\frac{4^{-j}}2\right)
  \zeta(2j)\zeta(2j+2)
 \sim-\frac{(2j)!}{j\pi^{4j+1}}\,.
 \label{sm:factorial}
\ee
The factors following the factorial approach one. The genus series in $1/N_f^2$, after removal of its leading $N_f^2$ term, consequently has zero radius of convergence. The binomial part converges for $|1/(3N_f^2\lamh)|<1$ and cannot cancel this factorial growth.

\begin{figure}[tbp]
 \centering\includegraphics[width=.70\textwidth]{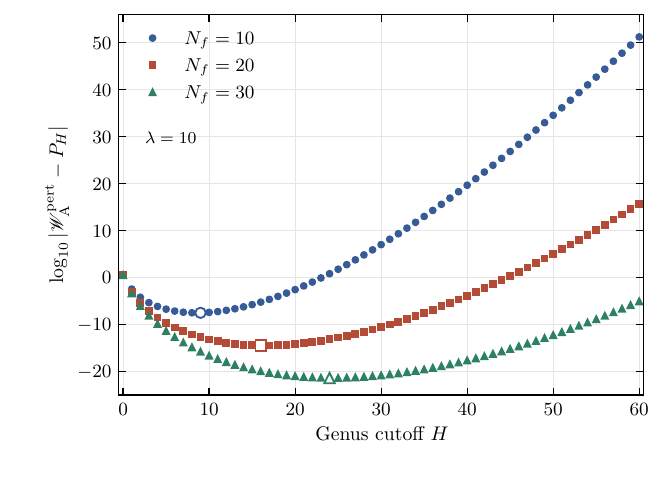}
 \caption{Genus truncation of the perturbative parent at $\lambda=10$. The ordinate is the logarithm of the absolute difference between Eq.~\eqref{sm:genusparent} and Eq.~\eqref{eq:genus-partial}. At $N_f=10,20,30$, the minima occur at $H=9,16,24$. Distinct markers identify the three flavor numbers; open markers indicate the minima. The comparison tests the parent expansion, not the full fixed-coupling Bethe observable.}
 \label{sm:genusfigure}
\end{figure}

For a quantitative illustration, define the finite partial sum
\be
 P_H(N,N_f)=-\sum_{h=0}^{H}
       (2\pi\ii\lambda)^{2h-2}\cW_h(\lambda)N^{2-2h}\,.
 \label{eq:genus-partial}
\ee
The symbol $P_H$ denotes this truncation only. Figure~\ref{sm:genusfigure} compares it with the unexpanded parent, evaluated using the finite Clausen constant. The finite-order minimum is a property of an asymptotic series and depends on the flavor number. It should not be confused with the convergent instanton truncations of Section~\ref{sec:modular}.

Finally, the fixed-flavor and fixed-coupling limits are different:
\be
 q=\exp\left[-2\pi\sqrt{2\left(\lamh+\frac1{3N_f^2}\right)}\right]
 \longrightarrow \exp\left[-2\pi\sqrt{2\lamh}\right]\ne0
 \quad(N_f\to\infty,\ \lambda\ {\rm fixed})\,.
 \label{eq:genus-nome-limit}
\ee
A correction exponentially small in $\sqrt N$ at fixed $N_f$ can therefore contribute to a genus coefficient at fixed $\lambda$. Such terms may carry worldsheet-instanton dependence, suppressed in a further large-$\lambda$ limit, even though they are not beyond all orders in $1/N_f$. The formulas above determine the genus expansion of the adopted perturbative parent. Extending them to the complete physical genus coefficients requires a generic-flavor remainder with control uniform in this limit; the three special-flavor completions do not supply it.

\section{Relations to ABJM and their physical scope}\label{sec:relations}

We write $\W_{\mathrm B}(M,k)=\im\Wt_{\rm ABJM}(M,k)$ for the critical value with the branch and additive convention of Refs.~\cite{Hosseini:2026dkj,Hosseini:2026lvj}. No ABJM vacuum sum is implicit in this notation. The comparisons below use the analytic completions proposed in those references.

\subsection{A common action and the mirror case}
We compare the closed expressions proposed for the two theories. The resulting functional identities follow exactly from these expressions; their identification with physical critical values is conditional on the respective completion conjectures. The ADHM expression in Eq.~\eqref{eq:completion} extends to real $N$ with $\Nh>0$ using the positive square root. The corresponding ABJM expression is~\cite{Hosseini:2026dkj,Hosseini:2026lvj}
\bea
 \W_{\mathrm B}(M,k)
 &=\frac{\pi^2\sqrt{2k}}6\widehat M^{3/2}+C_{\mathrm B}(k)
 +(1+t_{\mathrm B}D)\Phi_{\mathrm B,k}(\e^{-t_{\mathrm B}})\,,
 \\
 \widehat M&=M-\frac{k}{24}+\frac2{3k}\,,\qquad
 t_{\mathrm B}=2\pi\sqrt{2\widehat M/k}\,,
 \label{sm:ABcandidate}
\eea
where $D$ differentiates the primitive before its argument is set. For $k>0$ and $\widehat M>0$, we denote its perturbative parent by
\be
 \W_{\mathrm B}^{\mathrm{pert}}(M,k)
 =\frac{\pi^2\sqrt{2k}}6\widehat M^{3/2}+C_{\mathrm B}(k)\,.
 \label{eq:ABJM-parent}
\ee
For $k=1,2,4$, the constants are respectively
$-G-5\zeta(3)/(8\pi)$, $-5\zeta(3)/(2\pi)$, and $-3\zeta(3)/\pi$,
and the primitives are
\bea
 \Phi_{\mathrm B,1}&=\Phi_1\,,&\Phi_{\mathrm B,2}&=4\Phi_1\,,\\
 \Phi_{\mathrm B,4}(q)&=-\frac8\pi
                  [\cL_3(-q)-\tfrac14\cL_3(q^4)]\,.
 \label{sm:ABprimitive}
\eea
The positive square root specifies the continuation for $\widehat M>0$.
For $N_f=k=1$ their shifts, constants, and coefficient sequences coincide.
For ABJM level two, $\widehat M(2N+1,2)=2(N+5/8)$; its leading power
and tail are four times the flavor-one expressions, whereas
$C(1)-C_{\mathrm B}(2)/4=-G$. This proves
\be
 \boxed{
 \W_{\mathrm A}(N,1)
 =\W_{\mathrm B}(N,1)
 =\frac14\W_{\mathrm B}(2N+1,2)-G\,.
 \label{sm:integerrelation}
}
\ee

\subsection{Finite dissection identities}
For the finite relations at $N_f=2,4$, separating even and odd terms in
$\Li_3$ and then in its Lambert sum proves the standard dissection identity~\cite[Eq.~(2.52)]{Dorigoni:2020oon}
\begin{equation}
 \cL_3(-q)=-\cL_3(q)+\frac94\cL_3(q^2)-\frac14\cL_3(q^4)\,.
 \label{sm:dissection}
\end{equation}
Let $Sg(q)=g(q^2)$. In units of $1/\pi$, the primitive polynomials
acting on $\cL_3(q)$ are
\bea
 p_{\mathrm B,2}&=10-\frac{35}{2}S+\frac52S^2\,,\\
 p_{\mathrm B,4}&=8-18S+4S^2\,,\\
 p_{\mathrm A,2}&=-2+\frac12S^2\,,\\
 p_{\mathrm A,4}&=4-\frac72S-\frac74S^2+\frac12S^3\,.
 \label{sm:polynomials}
\eea
Direct multiplication gives
\bea
 p_{\mathrm A,2}
 &=\left(-\frac{11}{10}+\frac15S\right)p_{\mathrm B,2}
  +\left(\frac98-\frac18S\right)p_{\mathrm B,4}\,,\\
 p_{\mathrm A,4}
 &=\left(\frac{11}{10}-\frac15S\right)p_{\mathrm B,2}
  +\left(-\frac78+\frac14S\right)p_{\mathrm B,4}\,.
 \label{sm:polyidentities}
\eea
The chain rule gives
$(1+tD_q)\Phi(q^d)=[(1+dtD_z)\Phi(z)]_{z=q^d}$.
Thus each $S$ in Eq.~\eqref{sm:polyidentities} becomes $t\mapsto2t$,
or $\widehat M/k\mapsto4\widehat M/k$, for the tail.
The leading powers and constants are checked separately: doubling the
action multiplies the shifted $3/2$-power by $8$, and the ABJM zeta constants
give $-3\zeta(3)/(4\pi)$ and $-3\zeta(3)/(8\pi)$.
The remaining offsets are $-2G$ and $-7G$. At action $dt$, the ABJM rank is
\begin{equation}
 M_{k,d}=\frac{kd^2}{N_f}\left(N+\frac{7N_f}{24}+\frac1{3N_f}\right)
                  +\frac{k}{24}-\frac2{3k}\,.
 \label{sm:rankmap}
\end{equation}
Including the shifted powers and constants, the functional identities become
\be
\begin{aligned}
\W_{\mathrm A}(N,2)
={}&-\frac{11}{10}\W_{\mathrm B}(M_{2,1},2)
    +\frac15\W_{\mathrm B}(M_{2,2},2) \\
&+\frac98\W_{\mathrm B}(M_{4,1},4)
    -\frac18\W_{\mathrm B}(M_{4,2},4)
    -2G\,,
\\
\W_{\mathrm A}(N,4)
={}&\frac{11}{10}\W_{\mathrm B}(M_{2,1},2)
    -\frac15\W_{\mathrm B}(M_{2,2},2) \\
&-\frac78\W_{\mathrm B}(M_{4,1},4)
    +\frac14\W_{\mathrm B}(M_{4,2},4)
    -7G\,.
\end{aligned}
\label{sm:fullrelation}
\ee
For clarity, the four ABJM ranks in each relation are summarized in Table~\ref{tab:rankmap}.
\begin{table}[htp!]
\centering\small
\begin{tabular}{ccccc}
\toprule
$N_f$ & $M_{2,1}$ & $M_{2,2}$ & $M_{4,1}$ & $M_{4,2}$ \\
\midrule
$2$
& $N+\frac12$
& $4N+\frac{11}{4}$
& $2N+\frac32$
& $8N+6$ \\[2pt]
$4$
& $\frac N2+\frac38$
& $2N+\frac94$
& $N+\frac54$
& $4N+5$ \\
\bottomrule
\end{tabular}
\caption{ABJM ranks appearing in the functional identities
\eqref{sm:fullrelation}.}
\label{tab:rankmap}
\end{table}
The ranks $M_{k,d}$ in each equation use its own $N_f$ in
Eq.~\eqref{sm:rankmap}. These are identities of the full defined
functions, including their perturbative powers and constants.
Most such arguments are noninteger. These are relations between analytic expressions for critical values under the completion assumptions stated above, and do not define fractional-rank gauge theories or establish new dualities at flavors two and four.

\subsection{Negative nomes and level-four dilation identities}
The action dependence makes a negative-nome substitution more restrictive than an identity of primitives. Write $\W_{{\rm np},{\rm A},N_f}(t)$ and $\W_{{\rm np},{\rm B},k}(t)$ for the respective tails of the proposed completions as functions of their own action. At the primitive level,
$\Phi_{\mathrm B,4}(-q)=4\Phi_2(q)$.
Continuing $t$ along $t-\ii\theta$, $0\leq\theta\leq\pi$, gives $-q=\e^{-(t-\ii\pi)}$ and hence
\be
 \W_{{\rm np},{\rm B},4}(t-\ii\pi)
 =4\W_{{\rm np},{\rm A},2}(t)-4\ii\pi D\Phi_2(q)\,.
 \label{eq:negative_nome_correction}
\ee
The square root defining the continued rank is followed along the same path. The final term comes from the explicit multiplier $1+tD$ and cannot be discarded. The real-nome relations in Eq.~\eqref{sm:fullrelation} require no such continuation.

The level-two terms can instead be eliminated by a convergent infinite sum of level-four tails. With the dilation $Sg(q)=g(q^2)$ used above,
\bea
\frac{p_{\mathrm A,2}(S)}{p_{\mathrm B,4}(S)}
&=-\frac14\frac{1-S^2/4}{1-9S/4+S^2/2}
=\frac18-\frac{15}{56}\frac1{1-2S}
-\frac3{28}\frac1{1-S/4}\,,
\\
\frac{p_{\mathrm A,4}(S)}{p_{\mathrm B,4}(S)}
&=\frac14+\frac18S
-\frac{p_{\mathrm A,2}(S)}{p_{\mathrm B,4}(S)}\,.
\label{eq:rational-dilation}
\eea
These are identities of rational expressions in an operator, not pointwise quotients of Lambert sums. Set
\be
 d_j=\frac18\delta_{j0}-\frac{15}{56}2^j-\frac3{28}4^{-j}\,,
 \qquad j=0,1,\ldots\,,
\ee
where $\delta_{ab}$ is the Kronecker symbol. Finite geometric expansion, followed by the chain rule for the tail, gives
\bea
 \W_{{\rm np},{\rm A},2}(t)
 &=\sum_{j\geq0}d_j\W_{{\rm np},{\rm B},4}(2^jt)\,,\\
 \W_{{\rm np},{\rm A},4}(t)
 &=\sum_{j\geq0}
 \left(\frac14\delta_{j0}+\frac18\delta_{j1}-d_j\right)
 \W_{{\rm np},{\rm B},4}(2^jt)\,.
 \label{eq:level_four_dilation}
\eea
The coefficient in the second line follows directly from the second
identity in Eq.~\eqref{eq:rational-dilation}.
For fixed $t>0$, the tails at action $2^jt$ are
$O((2^jt+1)\e^{-2^jt})$. This estimate makes the sums absolutely convergent and sends the remainders of the finite geometric expansions to zero. Replacing the tails by the full expressions would introduce growing cubic powers and constants with divergent dilation sums. The perturbative subtraction is essential to Eq.~\eqref{eq:level_four_dilation}.

\subsection{Same-rank and affine-rank exclusions}

The shifted ranks already obstruct a naive replacement $k=N_f$ at the same physical rank. Let $s_{\rm A}=7N_f/24+1/(3N_f)$ and $s_{\rm B}=-k/24+2/(3k)$. Matching the $N^{3/2}$ and $N^{1/2}$ powers in a proportionality with an additive constant forces $s_{\rm A}=s_{\rm B}$. For the physical integer parameters,
\be
 s_{\rm A}-\frac58=\frac{(N_f-1)(7N_f-8)}{24N_f}\geq0\,,\qquad
 \frac58-s_{\rm B}=\frac{(k-1)(k+16)}{24k}\geq0\,.
 \label{eq:equal-rank-exclusion}
\ee
Both are saturated only at $N_f=k=1$.

Allowing a rank rescaling gives the more general ansatz
\be
 \W_{\rm A}(N,N_f)=a\W_{\rm B}(rN+b,k)+d\,,
 \qquad r>0\,,\quad k\in\{1,2,4\}\,,\quad N_f\in\{1,2,4\}\,,\label{eq:affineansatz}
\ee
with $a,b,d,r$ independent of $N$. Matching the leading and next algebraic powers forces
\be
 b=rs_{\rm A}-s_{\rm B}\,,\qquad a=\sqrt{N_f/k}\,r^{-3/2}\,.
\ee
The first nonzero exponential terms have actions $t$ and $\sqrt{rN_f/k}\,t$. Equality requires $r=k/N_f$, so $a=N_f^2/k^2$. In the common normalization $N_f^2/(2\pi)$, the first ADHM coefficients are $5,-1,1/2$ for flavors $1,2,4$, whereas the ABJM coefficients are $5,5,1$ at levels $1,2,4$. Only the pairs $(N_f,k)=(1,1),(1,2)$ survive. Their complete sequences coincide, as in Eq.~\eqref{sm:integerrelation}. These two relations therefore exhaust the single affine ansatz among the nine special pairs. This argument does not exclude nonlinear transformations, larger linear combinations, other levels, or a physical duality formulated in different observables.

For arbitrary $N_f$ and $k>0$, the action-matched map
\be
 M=\frac{k}{N_f}N+\frac{k}{3}+\frac{k}{3N_f^2}-\frac2{3k}
\ee
still gives the parent identity
\be
 \W_{\rm A}^{\rm pert}(N,N_f)
 =\frac{N_f^2}{k^2}\W_{\rm B}^{\rm pert}(M,k)
       +C(N_f)-\frac{N_f^2}{k^2}C_{\rm B}(k)\,.
 \label{eq:general-parent-map}
\ee
It follows by substituting $\widehat M=(k/N_f)\Nh$ into the leading power. The first exponential coefficient is independent information, and the special-flavor exclusions show that matching the parent alone is insufficient.

\section{High-precision validation}\label{sec:validation}
\subsection{The Bethe grid and numerical accuracy}
The calculation covers every integer $2\leq N\leq300$ and $1\leq N_f\leq30$, giving $8,970$ cases. For each fixed flavor, neighboring-rank continuation follows the ordered imaginary solution of Proposition~\ref{prop:branch}. Reflection reduces the number of unknowns; the first rank uses a triangular-density quantile seed, and subsequent ranks use interpolation and rescaling of the previous solution. Damped Newton steps preserve ordering. The numerical equations and potential evaluator contain none of the divisor coefficients in Eq.~\eqref{eq:divisor}.

Solver termination is followed by a separate evaluation of both the real logarithmic equations and the exponentiated complex product. An analytic residual-to-root bound then encloses the unique real solution, and the positive-decay potential is evaluated on enlarged root balls. The argument is given in Appendix~\ref{app:certificate}; the precision schedule and continuation parameters are in Appendix~\ref{app:numerics}. Arb/Acb arithmetic accounts for outward rounding in these production enclosures~\cite{Johansson:2017,PythonFlint:2026}. A small residual alone is not used as a root-error estimate.

The target precision is $800$ decimal digits for $N\leq100$ and $200$ for $N>100$. All cases satisfy the acceptance criteria, with minimum relative accuracies of $832$ and $224$ digits for nonzero roots and critical values in the two groups. Exact central roots at odd rank are excluded from relative-error counts. These are conservative minimum accuracies obtained from the root and potential enclosures, distinct from the working precision.

\subsection{Tests of the perturbative parent and constants}
The subtraction shown in Figure~\ref{fig:perturbative} uses Eq.~\eqref{eq:perturbative} with no fitted parameter. Separate fits of a shifted $3/2$-power plus a constant use windows $100$--$200$, $150$--$250$, and $200$--$300$. Both a fixed and a free leading coefficient were tested at $120$ digits. Repetitions at $170$ digits for $N_f=1,15,30$, both fit types, and all three windows preserve the stored fit parameters to $90$ absolute decimal places.

In the last window, the maximum absolute shift deviations are $3.78721\times10^{-9}$ with the leading coefficient fixed and $4.46873\times10^{-8}$ with it free, both at $N_f=30$. These are finite-window biases, not root errors or statistical confidence intervals. For comparison, at $N_f=30$ and $200\leq N\leq300$, fixing the shift to zero or to minus its predicted value and fitting only a constant gives maximum residuals about $43.35$ or $87.54$ in the normalization $\W_{\rm A}/(2\pi)$; the predicted shift gives $3.96\times10^{-8}$. This rejects those two alternatives on that grid, not every other perturbative ansatz.

\begin{figure}[tbp]
 \centering\includegraphics[width=.70\textwidth]{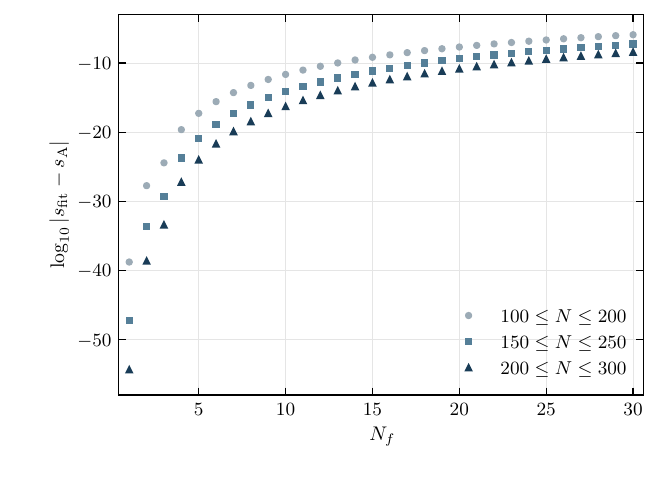}
 \caption{Shift fits with the leading coefficient and exponent $3/2$ fixed, and the additive shift and constant free. Moving the fit window upward reduces the deviation from $s_{\rm A}=7N_f/24+1/(3N_f)$. The vertical coordinate measures finite-rank contamination, not statistical uncertainty.}\label{sm:shiftfigure}
\end{figure}

The constant integral was evaluated at $125$ and $165$ decimal digits. The finite Clausen expressions were evaluated with ball arithmetic at $950$ digits and checked at $1,000$ digits. The exported constants have at least $926$ certified absolute decimal digits, including conversion to their stored decimal representations. Across all thirty flavors, the difference between $C(N_f)/(2\pi)$ from the finite sums and the $165$-digit quadrature is smaller than $4.80\times10^{-164}$. This quadrature comparison supplies empirical convergence evidence, while the finite-sum bounds enclose the analytically defined constant. The identification of that constant with the physical plateau is tested separately against the Bethe critical values.

\subsection{Whole-function comparisons}
Throughout this comparison, $\W_{\rm A}$ on the series side denotes the expression in Conjecture~\ref{conj:completion}, and $\W_{\rm A,M}$ is its truncation at $m=M$. Let $\bar w$ be the critical-value midpoint computed directly from the Bethe equations and $\eta$ its certified error allowance. For an interval $I_M$ enclosing the truncated series and an omitted-tail bound $\rho_M$, the interval comparison tests
\be
 I_{\infty}=I_M+[-\rho_M,\rho_M]\,,\qquad
 \sup_{z\in I_{\infty}}|\bar w-z|<\eta\,.\label{sm:comparison}
\ee
This is stronger than mere overlap of the two enclosures, but still allows a nonzero physical difference. The bound on the omitted tail follows from Eq.~\eqref{sm:tailbound}; the source allowance $\eta$ requires the separate root certificate. The convergent expressions are evaluated with ball arithmetic at $900$ decimal digits for $N_f=1$ and $950$ digits for $N_f=2,4$. At flavor one, the omitted-tail bound is smaller than $10^{-15}\eta$; at flavors two and four it is smaller than $10^{-910}$. The finite-sum enclosure is enlarged by this bound before comparison. All $897$ special-flavor cases satisfy Eq.~\eqref{sm:comparison}. Table~\ref{tab:candidates} reports the full-series difference bound $D_\infty=\sup_{z\in I_\infty}|\bar w-z|$, including truncation and numerical evaluation errors. No parameter is fitted in these comparisons.

\begin{table}[tbp]
 \centering\small
 \begin{tabular}{lrrr}
 \toprule
 Grid & Cases & $\max D_\infty$ & $\max(D_\infty/\eta)$\\
 \midrule
 $2\leq N\leq100$ &297&$2.055\times10^{-839}$&$1.728\times10^{-9}$\\
 $101\leq N\leq300$&600&$9.117\times10^{-231}$&$6.558\times10^{-10}$\\
 \bottomrule
 \end{tabular}
 \caption{Outward bounds on the full-series discrepancies in the $900$-digit ($N_f=1$) and $950$-digit ($N_f=2,4$) comparisons. The maxima are shown to four significant digits. Maxima in different columns need not occur at the same case. The largest difference bounds occur at $(N,N_f)=(99,1)$ and $(300,1)$. The error allowance $\eta$ comes from the Bethe-root enclosure.}\label{tab:candidates}
\end{table}

The exact $N=1$ benchmark $2(N_f+1)G$ agrees numerically with the expressions for all three special flavors; this numerical agreement is not an analytic summation proof at $N=1$. A distinct Lambert-index implementation, summing polylogarithms instead of divisor coefficients, reproduces eight cases at $N_f=2,4$ and $N=2,7,40,100$ at a $185$-digit comparison threshold using $200$-digit arithmetic. This is an implementation cross-check of the same series representation.

\begin{figure}[tbp]
 \centering\includegraphics[width=\textwidth]{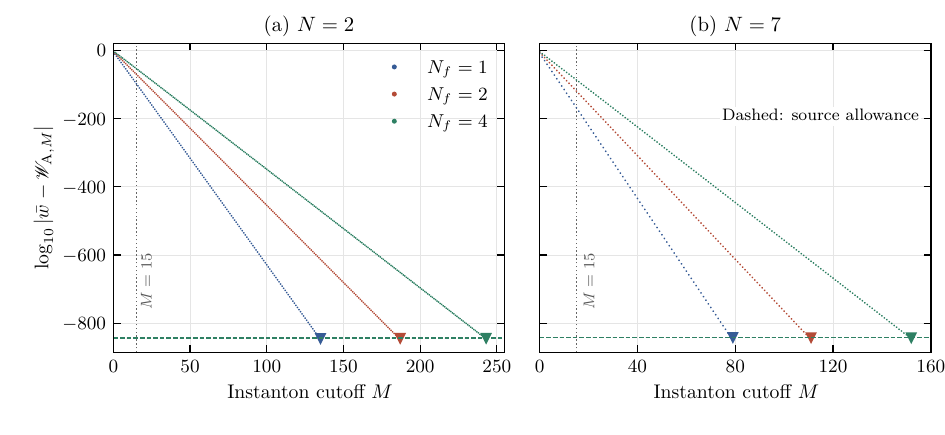}
 \caption{Truncation of the conjectured series at two ranks outside the specified ADHM reconstruction windows. Small filled markers show outward bounds on finite-sum discrepancies at the resolved integer cutoffs, without connecting lines. Downward triangles mark the first unresolved cutoffs at the critical-value error allowances, shown by dashed lines. The first unresolved cutoffs are $M=135,187,243$ at $N=2$ and $M=79,111,152$ at $N=7$, for $N_f=1,2,4$. The dotted line is $M=15$, the last rationally recovered sector at $N_f=2,4$; the fits contain $K=16$ sectors. The flavor-one completion is inherited from ABJM. No independent physical information is resolved below the source allowance. Agreement beyond the recovered sectors tests the continued function as a whole; it is not independent extraction of every coefficient up to the terminal cutoff.}\label{fig:convergence}
\end{figure}

The low-rank data therefore test far more of the proposed function than the finite fit can resolve sector by sector. The high-rank cases also extend outside the final fit windows. These ranks do not enter the coefficient fits specified in Section~\ref{sec:reconstruction}.

\subsection{Numerical resolution and the first omitted sector}
The numerical resolution in Figure~\ref{fig:convergence} is set by the critical-value enclosure. For the finite-series interval $I_M$, define
\be
 d_M^-=\inf_{z\in I_M}|\bar w-z|\,,\qquad
 d_M^+=\sup_{z\in I_M}|\bar w-z|\,.
 \label{eq:finite-discrepancy}
\ee
The first unresolved cutoff is the smallest $M$ for which $d_M^-\leq\eta$. At a cutoff satisfying this condition, the computed critical value cannot resolve a nonzero discrepancy from the finite sum. The plotted upper bound $d_M^+$ excludes the omitted infinite tail; Eq.~\eqref{sm:comparison} adds $\rho_M$ separately when comparing the full series.

The first omitted sector after a cutoff $M$ has the value
$N_f^2c_{M+1}^{(N_f)}(t+1/(M+1))q^{M+1}/(2\pi)$.
Its exponential scale accounts for the decrease with $M$, while cancellations and source uncertainty affect the final resolved point. No discrepancy below $\eta$ is interpreted as a newly measured physical correction. Likewise, reaching $M=243$ at one rank is a test of the continued function assembled from the whole sequence, not an independent extraction of $243$ coefficients.

The root and potential enclosures control numerical uncertainty in these comparisons, while Eq.~\eqref{sm:tailbound} controls truncation of the proposed series. The branch theorem, constant identities and convergence proof have their separate analytic basis. Equality of the series with the physical critical value remains Conjecture~\ref{conj:completion}.

\section{Discussion and outlook}\label{sec:discussion}
The finite-rank Bethe data support a weight-four organization of
nonperturbative corrections that extends beyond the
$N_f=k=1$ mirror pair. For $N_f=2,4$, the proposed ADHM
completions retain the Eisenstein-integral structure found in
ABJM, with divisor laws and minimal modular levels that
distinguish the theories. Their arithmetic cannot be recovered
by identifying flavor with Chern--Simons level. Resolving these
differences requires information beyond the leading continuum
density, including the additive constant whose subtraction
isolates the exponential remainder. The finite evaluations of
$C(N_f)$ fix this normalization, and the divisor sums organize
the proposed remainder into an absolutely convergent series at
every physical rank. The same convergent constant integral
generates factorially growing coefficients when expanded at
large flavor. It is this expansion that makes the type-IIA
genus series of the perturbative parent asymptotic, without
affecting the convergence of the special-flavor exponential
series.

Strict convexity and coercivity of the real variational potential
establish a unique solution of the prescribed logarithmic
equations for every integer $N\geq1$ and real $N_f>0$.
The symmetric chiral normalization makes the critical twisted
superpotential purely imaginary on the prescribed universal branch,
so the finite-rank comparison involves a uniquely specified
observable. The normalization of the proposed perturbative
contribution is determined independently by the integral
exchange identity, which fixes $A(1/2)$ and $a_2$ without
imposing equality of physical mirror constants. The Clausen
and multiple-polylogarithm expressions then evaluate the
integral-defined $C(N_f)$ at every positive integer flavor.
For $N_f=1,2,4$, the proposed exponential remainder also admits
analytic control. Cusp constants establish the minimal levels
of its Eisenstein generators, while the divisor coefficients
give unit convergence radii for the power-series components
and explicit tail estimates. These estimates bound the error
in evaluating the proposed completion at each physical rank.
Its identification with the Bethe critical value remains the
assertion of Conjecture~\ref{conj:completion}.

With the constants and divisor laws fixed, the proposed
completions agree with all $897$ special-flavor Bethe critical
values within their numerical error allowances. The critical
values are computed independently of the divisor ansatz, with
uncertainties controlled by residual-to-root bounds and separate
potential evaluations. The full-series comparisons also include
the analytic bounds on the omitted tails. At $N_f=2,4$, the
low-rank tests lie outside the fitting windows and require many
more than sixteen sectors to reach the available precision.
They thus probe the proposed continuation beyond the range
resolved by the coefficient fits. This agreement does not assign
the same accuracy to individual coefficients, whose least
resolved values remain sensitive to the rank window and sector
cutoff even when stable under increased arithmetic precision.
Equation~\eqref{eq:countercontinuation} exhibits the remaining
freedom through analytic deformations that preserve the first
sixteen coefficients and every tested value while changing
higher sectors. Selecting the physical completion from these
alternatives requires further constraints from the Bethe
problem.

\subsection{Generic flavor and the limits of the special pattern}
The generic-flavor Bethe data suggest that the leading
exponential correction retains the action $t$, with a prefactor
that is no longer proportional to $t+1$. The finite evaluations
of $C(N_f)$ fix the perturbative subtraction at every flavor number.
For $N_f\geq3$, the resulting remainder
is consistent with the empirical leading-sector approximation
\bea
 \W_A^{\rm np}(N,N_f)
 &\simeq
 \frac{N_f^2}{2\pi}
 \bigl[A_1(N_f)t+B_1(N_f)\bigr]\e^{-t}\,,\\
 A_1(N_f)
 &=\frac12+\cos\frac{2\pi}{N_f}\,,\\
 B_1(N_f)
 &=A_1(N_f)-\frac{\pi}{N_f}\cot\frac{2\pi}{N_f}
       \left(3+4\cos\frac{2\pi}{N_f}\right)\,.
 \label{eq:generic-sector}
\eea
At $N_f=4$, the coefficients reduce to $A_1=B_1=1/2$,
reproducing the first sector of
Conjecture~\ref{conj:completion}. At $N_f=3$, the formula gives
$A_1=0$ and $B_1=\pi/(3\sqrt3)$, predicting a nonzero
$\e^{-t}$ correction with no $t\e^{-t}$ term.

We test Eq.~\eqref{eq:generic-sector} at $N=100,200,300$
and $N_f=3,\ldots,30$ using $250$-digit arithmetic.
The absolute discrepancy between the measured
$\W_A^{\rm np}$ and the proposed first sector, divided by
$|\W_A^{\rm np}|$, has maxima $3.83\times10^{-6}$,
$8.14\times10^{-9}$ and $6.35\times10^{-11}$ over the
tested flavors at the three ranks, respectively. All three
maxima occur at $N_f=30$. Their decrease with rank is consistent
with further exponential suppression, although the comparison
does not determine the higher-sector coefficients.

If Eq.~\eqref{eq:generic-sector} gives the correct leading
sector, the generic inequality $A_1\neq B_1$ requires a
modification of the special $(t+1/m)$ prefactor.
Changing the divisor coefficients alone cannot accommodate
this dependence, since the special pattern enforces
$B_m=A_m/m$ in every sector. Whether a completion with the
more general prefactor admits a modular organization remains
open. The cases $N_f=1,2,4$ therefore do not constitute a
classification of flavors admitting closed or modular
expressions.

\subsection{Holographic interpretation and the missing physical derivation}
At $N_f=k=1$, the rank shifts, constants and divisor laws of
the two proposed completions coincide, consistently with mirror
symmetry. For $N_f=2,4$, the dissection identities of
Section~\ref{sec:relations} express the ADHM functions as finite
linear combinations of level-two and level-four ABJM functions,
together with explicit additive constants. These identities
include the shifted powers and the entire proposed exponential
remainder. The rescalings of the instanton action fix the ABJM
rank arguments, which are generally noninteger, so the relations
concern analytic continuations of the proposed critical values
and do not establish dualities between gauge theories. For these
two flavors, the first exponential coefficients exclude a
reduction to a single affine-rank ABJM expression within the
tested special-level class, even after the shifted powers and
actions have been matched. This exclusion concerns equal-rank
ABJM Bethe observables and does not constrain the exact
round-sphere identity relating two-flavor ADHM to the normalized
magnitude of the $\U(N)_2\times\U(N+1)_{-2}$ ABJ partition
function~\cite[Eq.~(2.2)]{Hatsuda:2014vsa}.

The finite-rank Cardy relation~\eqref{eq:intro-cardy} transfers
this flavor-dependent arithmetic to the leading exponent of
the selected vacuum contribution to the superconformal
index~\cite{Choi:2019dfu,Bobev:2022wem,Bobev:2024mqw}.
For $N_f=1,2,4$, substituting Conjecture~\ref{conj:completion}
gives
\be
 \log\mathcal I_{\mathrm{vac}}(\omega;N,N_f)
 =-\frac{1}{\pi\omega}
 \left[\frac{\pi^2\sqrt{2N_f}}6\Nh^{3/2}+C(N_f)
 +\frac{N_f^2}{2\pi}\sum_{m\geq1}c_m^{(N_f)}
       \left(t+\frac1m\right)q^m\right]+O(\omega^0)\,.
 \label{eq:cardy_completed}
\ee
At fixed $N,N_f$, the nome $q$ is independent of $\omega$,
so every exponential sector contributes to the $1/\omega$
coefficient as $\omega\to\ii0^+$. The Hessian and the remaining
one-loop factors enter the $O(\omega^0)$ term. In a chamber
where the selected saddle contributes to the integration cycle
and dominates the competing vacua, Eq.~\eqref{eq:cardy_completed}
also fixes the leading exponent of the full index.
Establishing this dominance requires information beyond the
uniqueness of the real logarithmic solution. The leading
large-$N$ critical value supplies the field-theory input to
gravitational blocks whose gluing yields supersymmetric
$\mathrm{AdS}_4$ black hole entropy
functions~\cite{Hosseini:2019iad}.
The constant and divisor sectors in
Eq.~\eqref{eq:cardy_completed} specify the additional
contributions that a quantum extension of this construction
would have to reproduce in the Cardy limit.

Fixing the boundary value of the M-theory three-form potential $A_3$
defines a grand-canonical ensemble in asymptotically $\mathrm{AdS}_4$
holography, with chemical potential $\mu$ conjugate to the rank
$N$~\cite{Gautason:2025plx}. Relative equivariant localization
expresses the cubic and linear $\mu$-dependence of the
supersymmetric action in terms of the eleven-dimensional
Chern--Simons couplings, including the $A_3\wedge X_8$
curvature correction~\cite{BenettiGenolini:2026cyc}.
The all-order extension assumes an equivariant completion of
the quantum action and cancellation of the relevant boundary
contributions. An inverse Laplace transform converts the result
to fixed rank, with the linear term producing a shift of $N$
in the $N^{3/2}$ saddle exponent. For a flat measure, the full
transform gives an Airy function. The relation between the
measure and observable-dependent logarithmic corrections has
been analyzed for ABJM partition functions and
indices~\cite{Bobev:2026gir}. The proposed one-loop determinant
in Ref.~\cite{BenettiGenolini:2026kjy} also gives closed
expressions for the ABJM twisted-superpotential constant and the
constant in the logarithm of the topologically twisted index.
In the unrefined expansion, the simple-pole coefficient in
the logarithm of each zeta-regularized fixed-point contribution
determines the twisted-superpotential constant, reproducing the
ABJM combination in Eq.~\eqref{eq:exchange} at the universal
specialization. For the topologically twisted index, the
unrefined limit is taken in the combined fixed-rank expression
after evaluating the inverse Laplace integrals. The poles
cancel between the two fixed-point contributions, while their
logarithmic singularities cancel against the Gaussian terms.
The remaining rank-independent term gives the twisted-index
constant at general chemical potentials and magnetic fluxes.
These results depend on the proposed fluctuation character
and the chosen measure and integration cycles. An extension
to ADHM, with a fluctuation complex appropriate to the
superconformal-index saddle, should reproduce the shift
$s_{\rm A}=7N_f/24+1/(3N_f)$ and the contribution
$-C(N_f)/(\pi\omega)$ in Eq.~\eqref{eq:cardy_completed}.
The corresponding twisted-index constant would contribute
at order $\omega^0$. The proposed divisor sums require
nonperturbative bulk contributions beyond the localized
action and perturbative fluctuation determinant.

Wrapped Euclidean M2-branes give a semiclassical description of
nonperturbative corrections to ABJM observables. Localization of
their zero modes and one-loop quantization reproduce the leading
round-sphere instanton contribution, with a grand-canonical
interpretation, and yield predictions for other supersymmetric
boundary conditions~\cite{Gautason:2025per}. The ADHM
Lambert primitives suggest a multiple-cover organization of the
corresponding sectors. In the expansion
$\Li_3(q^r)=\sum_{s\geq1}q^{rs}/s^3$, the factorizations $m=rs$
generate the divisor sums, with inverse-cubic weights for the
putative covering number $s$. A bulk derivation would have to
identify the supersymmetric cycles and account for these weights
and the flavor-dependent combinations entering $\Phi_{N_f}$.
After transformation to fixed rank, the $1/\omega$ coefficient
of $\log\mathcal I_{\mathrm{vac}}$ should reproduce the full
$t+1/m$ dependence, together with the signs and divisibility
patterns of $c_m^{(N_f)}$. The flavor dependence also controls
how such sectors contribute in the type-IIA limit. As
$N,N_f\to\infty$ at fixed $\lambda=N/N_f$, the nome approaches
$e^{-2\pi\sqrt{2\lamh}}$, so exponential suppression in rank
becomes exponential suppression at strong coupling. These terms
have the scale expected of worldsheet instantons and can enter
the genus coefficients at ordinary orders in $1/N_f$.
Determining their contribution requires a generic-flavor
completion with uniform control in this joint limit. This would
supply the instanton dependence absent from the genus expansion
of the perturbative parent.

The proposed completion constrains the entire sequence of
nonperturbative Bethe data through a four-dimensional Eisenstein
space. Within the span of $E_4(q^d)$ with $d=1,2,4,8$, the Fourier
coefficients at $q,q^2,q^4,q^8$ determine the generating form
uniquely. The zero-mode prescription fixes its Eichler primitive,
and the operator $1+tD$ then determines every exponential sector.
This finite determination depends on the representation itself,
including the restriction to the specified Eisenstein span, which
has yet to be derived from the Bethe equations.
An analytic derivation of this representation and its independent
coefficients would establish the nonperturbative identification
in Conjecture~\ref{conj:completion}. A functional relation for the
critical value, obtained directly from the Bethe problem and
supplemented by sufficient branch and asymptotic data, could
determine the same completion without assuming a modular ansatz.
The recurrence of weight four in ADHM and ABJM beyond the mirror
pair suggests a common constraint on their nonperturbative Bethe
data. Understanding how such a constraint arises, and how its
solutions depend on the matter content, would give a microscopic
explanation of the different divisor laws entering their
finite-rank Cardy exponents.

\section*{Acknowledgments}
This research was supported by UK Research and Innovation (UKRI) under the UK government's Horizon Europe funding guarantee (Grant No.~EP/Y027604/1).

\textbf{Statement on use of AI.}
The ideas, arguments, and writing in this paper came from the author. An OpenAI model provided useful assistance with code development, computational checks, and copy editing.

\section*{Data and code availability}
The data and code supporting the findings of this manuscript are not publicly available and may be obtained from the author upon reasonable request.

\appendix
\section{A root-of-unity representation of the constant map}
\label{app:constants}

Setting $u=\e^{-x}$ in Eq.~\eqref{eq:constant} writes its kernel as
\be
 \frac12\log\cosh x+\frac14\log\cosh\frac x2
 =-\frac58\log u+\frac12\log(1+u^2)
                    +\frac14\log(1+u)-\frac34\log2\,.
 \label{eq:MPL_kernel}
\ee
The polynomials $1-u^{N_f}$ and $1+u^2$ factor into linear
terms at roots of unity. The resulting products of logarithms
have integrals expressed by depth-two multiple polylogarithms,
giving a finite representation of the flavor-dependent part
of $C(N_f)$ directly from the ADHM kernel.

For $|a|=|b|=1$, we use the convention
\be
 \Li_{2,1}(a,b)=\sum_{r>s\geq1}\frac{a^rb^s}{r^2s}\,,
 \qquad
 T(a,b)=\Li_{2,1}(a,b/a)+\Li_{2,1}(b,a/b)\,.
 \label{eq:MPL_definition}
\ee
These series converge absolutely, with each multiple
polylogarithm bounded in absolute value by
$\sum_{r\geq2}H_{r-1}/r^2<\infty$, where
$H_j=\sum_{h=1}^j1/h=O(\log(j+1))$.
Expanding the logarithms and integrating termwise gives
\be
 \int_0^1\log(1-au)\log(1-bu)\frac{\rd u}{u}
 =\sum_{h,\ell\geq1}\frac{a^hb^\ell}{h\ell(h+\ell)}
 =T(a,b)\,.
 \label{eq:MPL_log_integral}
\ee
Indeed, setting $r=h+\ell$ and using
\be
 \frac1{\ell(r-\ell)}
 =\frac1r\left(\frac1\ell+\frac1{r-\ell}\right)
\ee
separates the double sum into the two terms in
Eq.~\eqref{eq:MPL_definition}. The integrated absolute
double sum is $2\sum_{r\geq2}H_{r-1}/r^2$, which justifies
the interchange also at unit-modulus arguments.
For $0<u<1$, the factors $1-au$ and $1-bu$ lie in the
open right half-plane. Their principal logarithms are
therefore continuous along the integration path, and the
possible logarithmic singularities at $u=1$ are integrable.

Define the finite sum
\be
 M_{N_f}(a)
 =\sum_{j=0}^{N_f-1}T(a,\e^{2\pi\ii j/N_f})\,.
 \label{eq:MPL_root_sum}
\ee
The factorization
$1-u^{N_f}=\prod_{j=0}^{N_f-1}(1-\e^{2\pi\ii j/N_f}u)$
identifies the sum of the principal logarithms with
$\log(1-u^{N_f})$. Both expressions vanish at $u=0$
and remain continuous for $0\leq u<1$, fixing any possible
multiple of $2\pi\ii$. Equation~\eqref{eq:MPL_log_integral}
then gives
\be
 M_{N_f}(a)=\int_0^1
       \log(1-u^{N_f})\log(1-au)\frac{\rd u}{u}\,.
 \label{eq:MPL_M_integral}
\ee

The remaining terms in the kernel are evaluated by the moments
\be
 \int_0^1\log u\log(1-u^{N_f})\frac{\rd u}{u}
 =\frac{\zeta(3)}{N_f^2}\,,
 \qquad
 \int_0^1\log(1-u^{N_f})\frac{\rd u}{u}
 =-\frac{\zeta(2)}{N_f}\,.
 \label{eq:MPL_moments}
\ee
They follow by expanding $\log(1-u^{N_f})$ into its
absolutely integrable power series. The constant term
$-\frac34\log2$ in Eq.~\eqref{eq:MPL_kernel} contributes
$+\pi\log2/4$ to $C(N_f)$, cancelling the explicit
$-\pi\log2/4$ in Eq.~\eqref{eq:constant}.
Using
$\log(1+u^2)=\log(1-\ii u)+\log(1+\ii u)$
and the value of $a_2$ in Eq.~\eqref{eq:a2}, we obtain
\be
 C(N_f)=N_f^2\left(\frac{7\zeta(3)}{32\pi}-\frac G2\right)
          -\frac{5\zeta(3)}{4\pi N_f}
 +\frac{N_f}{\pi}
  \left[M_{N_f}(\ii)+M_{N_f}(-\ii)
                    +\frac12M_{N_f}(-1)\right]\,.
 \label{eq:constant_MPL}
\ee
Equation~\eqref{eq:MPL_M_integral} implies
$M_{N_f}(-\ii)=\overline{M_{N_f}(\ii)}$ and
$M_{N_f}(-1)\in\mathbb R$, making the reality of this
expression explicit. All multiple-polylogarithm arguments
in Eq.~\eqref{eq:constant_MPL} are roots of unity whose
orders divide $\operatorname{lcm}(4,N_f)$.

The normalization involving $A(1/2)$ can also be expressed
through a fourth-root sum,
\be
 J_4=\sum_{a\in\{1,-1,\ii,-\ii\}}\Li_{2,1}(1,a)\,.
 \label{eq:J4_definition}
\ee
The series convention in Eq.~\eqref{eq:MPL_definition} gives
\be
 \Li_{2,1}(1,a)
 =\int_0^1\frac{\log u\log(1-au)}{1-u}\,\rd u\,.
\ee
To verify this identity, expand $(1-u)^{-1}$ and
$\log(1-au)$ and use
$\int_0^1u^{r-1}\log u\,\rd u=-1/r^2$.
Absolute convergence justifies termwise integration,
including at $a=1$. Summing over the four roots replaces
$\log(1-au)$ by $\log(1-u^4)$.
Consequently, the substitution $u=\e^{-x/2}$ in
Eq.~\eqref{sm:a2} at $k=1/2$ yields
\be
 \int_0^\infty
 \frac{x\log(1-\e^{-2x})}{\e^{x/2}-1}\,\rd x
 =-4\int_0^1\frac{\log u\log(1-u^4)}{1-u}\,\rd u
 =-4J_4\,,
\ee
and hence
\be
 A(1/2)=\frac{127\zeta(3)}{32\pi^2}-\frac{J_4}{\pi^2}\,.
 \label{eq:Ahalf_MPL}
\ee
Combining this identity with the analytic value of $A(1/2)$
established in Eq.~\eqref{eq:a2} evaluates the root sum,
\be
 J_4=4\zeta(3)-\frac{\pi G}{2}
                   -\frac{7\pi^2}{16}\log2\,.
 \label{eq:J4_evaluation}
\ee
Substitution into Eq.~\eqref{eq:a2definition} then gives
\be
 2\pi a_2
 =-\frac{121\zeta(3)}{32\pi}
   +\frac{7\pi}{16}\log2+\frac{J_4}{\pi}
 =\frac{7\zeta(3)}{32\pi}-\frac G2\,,
\ee
reproducing the coefficient of $N_f^2$ in
Eq.~\eqref{eq:constant_MPL}.

\section{A residual-to-root enclosure}
\label{app:certificate}

The monotonicity of the interactions in Eq.~\eqref{eq:bethe}
allows an independently bounded residual to control the error
in each Bethe root. On the faces of a cube centred at an
approximate solution, the pairwise terms have a definite sign
relative to their values at the centre. A lower bound on the
fundamental derivative then fixes a sufficient enclosure radius.

Let $\bar{\boldsymbol y}\in\mathbb R^N$ be an approximate
solution, and choose bounds satisfying
\bea
 \epsilon&\geq\max_j|B_j(\bar{\boldsymbol y})|\,,
 \qquad
 M_0\geq\max_j|\bar y_j|\,,
 \\
 \delta&\geq\frac{2\epsilon\cosh(M_0+1)}{N_f}\,,
 \qquad
 0<\delta<1\,.
 \label{sm:certificate}
\eea
If $\epsilon=0$, the approximate vector is already the exact
solution by Proposition~\ref{prop:branch}. For $\epsilon>0$,
consider the closed cube
\be
 \mathcal Q_\delta
 =\left\{\boldsymbol y\in\mathbb R^N:
 \|\boldsymbol y-\bar{\boldsymbol y}\|_\infty
 \leq\delta\right\}\,.
\ee
On its upper $j$-face, $y_j=\bar y_j+\delta$, while
$y_h\leq\bar y_h+\delta$ for every $h$. Hence
$y_j-y_h\geq\bar y_j-\bar y_h$, and monotonicity of $f$
bounds each pairwise contribution from below by its value
at $\bar{\boldsymbol y}$. Along the fundamental segment,
$|s|\leq M_0+\delta<M_0+1$, so
$f'(s)=\sech s\geq\sech(M_0+1)$. It follows that
\bea
 B_j(\boldsymbol y)
 &\geq B_j(\bar{\boldsymbol y})
       +N_f\bigl[f(\bar y_j+\delta)-f(\bar y_j)\bigr]
 \\
 &\geq-\epsilon+N_f\delta\sech(M_0+1)
 \geq\epsilon>0\,.
\eea
On the lower face, $y_j=\bar y_j-\delta$, the same argument
with reversed inequalities gives
$B_j(\boldsymbol y)\leq-\epsilon<0$.

The continuous variational function $\Psi$ attains a minimum
on $\mathcal Q_\delta$. Since $\nabla\Psi=\boldsymbol B$,
the face inequalities imply that an inward displacement
normal to any active face decreases $\Psi$. This excludes
a boundary minimum, including at intersections of faces.
The minimum therefore lies in the interior and satisfies
$\boldsymbol B=0$. Proposition~\ref{prop:branch} identifies
it with the unique real logarithmic solution, proving
\be
 \|\boldsymbol y^*-\bar{\boldsymbol y}\|_\infty
 \leq\delta\,.
\ee
The enclosure follows from the gradient structure and the
residual bound, independently of the convergence history
of the Newton iteration.

For a component satisfying $\delta<|\bar y_j|$, the triangle
inequality gives $|y_j^*|\geq|\bar y_j|-\delta>0$ and hence
\be
 \frac{|y_j^*-\bar y_j|}{|y_j^*|}
 \leq\frac{\delta}{|\bar y_j|-\delta}\,.
 \label{eq:relative-root-bound}
\ee
In numerical evaluation, the denominator is replaced by a
strictly positive downward-rounded lower bound, and the
quotient is bounded from above. Rounding the negative
base-ten logarithms of the absolute and relative error
bounds downwards gives lower bounds on the corresponding
decimal accuracies. At odd rank, reflection symmetry fixes
the central root exactly to zero, so no relative accuracy
is assigned to that component.

The residual and the quantities entering
Eq.~\eqref{sm:certificate} must themselves be evaluated with
outward bounds. The serialized decimal coordinates are
regarded as exact centres, with any conversion error included
in their numerical representation. A point-arithmetic
residual, even one rounded to zero, does not establish the
required inequality for the exact residual.

After verifying that the component intervals are disjoint
and ordered, interval evaluation of Eq.~\eqref{sm:potential}
encloses the critical value $\W_{\mathrm A}(N,N_f)$.
The enclosure includes the root uncertainty, rounding in
the transcendental functions, and any enlargement required
when exporting the result. It supplies the Bethe-side error
allowance independently of the perturbative subtraction
and the proposed divisor laws. Errors in fitted coefficients
and truncations of the conjectured series require separate
bounds.

\section{Numerical implementation}\label{app:numerics}
\subsection{Reduced equations and continuation}

Reflection symmetry in Proposition~\ref{prop:branch} reduces
the numerical problem to the positive roots. With
$m=\lfloor N/2\rfloor$ and $e=N\bmod2$, write
\[
 \boldsymbol y=(-s_m,\ldots,-s_1,[0],s_1,\ldots,s_m)\,,
 \qquad 0<s_1<\cdots<s_m\,,
\]
where the bracketed zero is present only at odd rank.
For $i=1,\ldots,m$, the positive-root components of
Eq.~\eqref{eq:bethe} become
\be
 R_i(\boldsymbol s)
 =(N_f+2e)f(s_i)+2f(2s_i)
 +2\sum_{j\ne i}\bigl[f(s_i-s_j)+f(s_i+s_j)\bigr]
 -\pi(2i+e-1)=0\,.
 \label{eq:reduced}
\ee
The reflected partner $-s_i$ contributes $2f(2s_i)$,
and the central root at odd rank contributes $2f(s_i)$.
Differentiation gives the reduced Jacobian
$H_{ij}=\partial R_i/\partial s_j$,
\bea
 H_{ii}
 &=(N_f+2e)\sech s_i+4\sech(2s_i)
   +2\sum_{j\ne i}
      \bigl[\sech(s_i-s_j)+\sech(s_i+s_j)\bigr]\,,\\
 H_{ij}
 &=2\bigl[\sech(s_i+s_j)-\sech(s_i-s_j)\bigr]\,,
 \qquad i\ne j\,.
 \label{eq:reduced-jacobian}
\eea

For each fixed $N_f$, continuation begins at $N=2$ with
the positive quantiles of the triangular density in
Eq.~\eqref{eq:continuum_density}. In the seeding formulas,
$k=N-m+i-1$ denotes the zero-based position of $s_i$ in
the full root vector. The initial guess is
\be
 T_N=\pi\sqrt{\frac{N}{2N_f}}\,,
 \qquad
 p_k=\frac{k+1/2}{N}\,,
 \qquad
 s_i^{\rm seed}
 =T_N\bigl[1-\sqrt{2(1-p_k)}\bigr]\,.
 \label{eq:initial-seed}
\ee
A related continuation scheme, combining eigenvalue interpolation,
$\sqrt N$ rescaling and damped Newton refinement, was used to
solve the $S^3$ saddle equations of chiral Chern--Simons--matter
theories in Ref.~\cite{Hosseini:2025jxb}.
For $N\geq3$, set $N_0=N-1$ and denote the previously
computed full root vector by
$(y_0^{\rm old},\ldots,y_{N_0-1}^{\rm old})$, again using
zero-based indices. Interpolation at the new quantiles,
followed by the leading $\sqrt N$ rescaling, gives
\bea
 v&=\frac{(k+1/2)N_0}{N}-\frac12\,,
 \qquad
 j=\max\bigl(0,\min(N_0-2,\lfloor v\rfloor)\bigr)\,,\\
 s_i^{\rm seed}
 &=\sqrt{\frac{N}{N_0}}
   \bigl[(1-v+j)y_j^{\rm old}
              +(v-j)y_{j+1}^{\rm old}\bigr]\,.
 \label{eq:continuation-seed}
\eea
The coordinate $v$ locates the new quantile on the old
root grid. Clipping $j$ selects adjacent old roots while
allowing linear extrapolation at the endpoints. This
predictor initializes a new solve of
Eq.~\eqref{eq:reduced} at each rank.

A Newton correction satisfies
$H\boldsymbol d=\boldsymbol R$. Starting with the full
step, successive halvings give trial vectors
$\boldsymbol s-2^{-h}\boldsymbol d$. A trial is accepted
only if its components remain positive and ordered and
the maximum residual $\|\boldsymbol R\|_\infty$ decreases.
There are at most $35$ Newton iterations and $30$ halvings
per precision stage, with iterates reset to their ball
midpoints.

For a target accuracy of $p$ decimal digits, the working
precision starts at $p_0=40$ and increases according to
$p_{r+1}=\min(p+45,2p_r-10)$. The residual threshold at
stage $r$ is $10^{-(p_r-12)}$.
Table~\ref{tab:precision} gives the resulting working
precisions and export lengths. Final acceptance requires
the independent residual checks described below and the
root and critical-value enclosures of
Appendix~\ref{app:certificate}.

\begin{table}[tbp]
\centering
\begin{tabular}{crrrrr}
\toprule
Target $p$&Cases&Solver&Verifier&Potential&Root/potential export\\
\midrule
800&2,970&845&900&870&855 / 870\\
200&6,000&245&300&270&255 / 270\\
\bottomrule
\end{tabular}
\caption{Decimal working precisions and exported significant digits. The solver stages are $40,70,130,250,490,845$ for $p=800$ and $40,70,130,245$ for $p=200$. Exported length is distinct from certified accuracy.}
\label{tab:precision}
\end{table}

\subsection{Independent residuals and acceptance}

The verifier reads the serialized decimal centres $\bar y_j$
as exact inputs and evaluates the full logarithmic
system~\eqref{eq:bethe} independently of the reduced Newton
solver. It also computes the complex residuals directly
from the exponentiated equations,
\be
 E_j(\bar{\boldsymbol y})
 =
 \left(
 \frac{1+\ii\e^{-\bar y_j}}
      {\e^{-\bar y_j}+\ii}
 \right)^{N_f}
 \prod_{h\ne j}
 \left(
 \frac{\e^{-\bar y_j}-\ii\e^{-\bar y_h}}
      {\e^{-\bar y_j}+\ii\e^{-\bar y_h}}
 \right)^2-1\,.
 \label{eq:complex-residual}
\ee
The identity
$E_j(\bar{\boldsymbol y})
=\e^{-\ii B_j(\bar{\boldsymbol y})}-1$
shows that the complex residual vanishes whenever
$B_j\in2\pi\mathbb Z$. The logarithmic check is therefore
needed to distinguish the prescribed branch, even when
the exponentiated equations are satisfied to high precision.

For a target of $p$ decimal digits, outward upper bounds
on both $\max_j|B_j|$ and $\max_j|E_j|$ must lie strictly
below $10^{-p}$. The logarithmic residual bound then enters
the enclosure of Appendix~\ref{app:certificate}, which
bounds the error in the roots independently of the Newton
iteration. Acceptance requires disjoint, ordered root
intervals and reflection symmetry. Evaluating
Eq.~\eqref{sm:potential} on these intervals encloses the
critical value, and both this value and every nonzero
root must have at least $p$ certified relative digits.
The central root at odd rank is fixed exactly to zero
and is excluded from the relative-accuracy requirement.
The working precisions are listed in
Table~\ref{tab:precision}. Different flavors are processed
separately because the arithmetic precision context is
process-wide.

All cases on the grid satisfy these criteria. The minimum
certified decimal accuracies in the two precision groups are
\be
 \begin{array}{c|rrrr}
 p
 &\text{root absolute}
 &\text{root relative}
 &\text{potential absolute}
 &\text{potential relative}\\
 \hline
 800&833&832&828&832\\
 200&225&224&219&224
 \end{array}\,.
 \label{eq:accuracy-minima}
\ee
These lower bounds follow from the production enclosures,
including allowances for numerical conversion and decimal
export. They measure the accuracy of the roots and critical
values, without assigning uncertainties to coefficients
subsequently inferred from fits. The full grid contains
$30\sum_{N=2}^{300}N=1{,}354{,}470$ roots.

\subsection{Fitting and truncation}

For the free-sector fits of Section~\ref{sec:reconstruction},
define
$\delta=(2\pi/N_f^2)(\W_{\rm A}-\W_{\rm A}^{\rm pert})$
and $q_0=q(N_{\min})$. The QR least-squares problem uses
the normalized target $\delta/q$ and the paired columns
\be
 \left(t(q/q_0)^{m-1},\ (q/q_0)^{m-1}\right)_{m=1}^{K}\,.
 \label{eq:qr-design}
\ee
The fitted entries are $q_0^{m-1}A_m$ and $q_0^{m-1}B_m$,
so division by $q_0^{m-1}$ recovers the coefficients of
Eq.~\eqref{sm:fit}. This invertible column rescaling changes
the conditioning while preserving the least-squares problem
for the target $\delta/q$ in exact arithmetic.
All $2K$ coefficients remain free. Neither the divisor laws
nor the relation $B_m=A_m/m$ enters the fit or the subsequent
rational reconstruction.

For each coefficient, let $\alpha_{10}$ and $\alpha_{20}$
denote the estimates from the $10$--$100$ and $20$--$100$
fitting windows. Their decimal exports retain $200$
significant digits and are parsed as exact rationals.
The difference between the estimates and their export
resolution define
\bea
 \epsilon_\alpha
 &=\max\left\{|\alpha_{10}-\alpha_{20}|,
 10^{\lfloor\log_{10}|\alpha_{10}|\rfloor-199},
 10^{\lfloor\log_{10}|\alpha_{20}|\rfloor-199}\right\}\,,\\
 I_\alpha
 &=\left[\min(\alpha_{10},\alpha_{20})-10\epsilon_\alpha,
         \max(\alpha_{10},\alpha_{20})+10\epsilon_\alpha
   \right]\,.
 \label{eq:rational-window}
\eea
The two powers of ten correspond to one unit in the last
exported digit and prevent the interval from collapsing
when the rounded estimates coincide. The factor of ten
specifies the empirical margin used to assess window
sensitivity.

Continued fractions select the nearest reduced rational
with denominator at most $10^6$ to each of $\alpha_{10}$,
$\alpha_{20}$ and their midpoint. Recovery requires all
three selections to coincide at a fraction $r/s\in I_\alpha$.
Its immediate lower and upper neighbors among reduced
fractions in the same denominator class must lie strictly
below and above $I_\alpha$, respectively. These conditions
are checked in exact integer arithmetic.

For the lower neighbor $r_-/s_-$, the determinant relation
is $rs_--r_-s=1$, with $s_-$ the largest denominator in
its congruence class satisfying $s_-\leq10^6$.
Consequently $s+s_->10^6$. Any fraction $v/w$ strictly
between $r_-/s_-$ and $r/s$ obeys
\be
 w=s(s_-v-r_-w)+s_-(rw-sv)
   \geq s+s_->10^6\,,
 \label{eq:farey-uniqueness}
\ee
since both parenthesized terms are positive integers.
The upper-neighbor argument gives the same denominator
bound. The test therefore proves that $r/s$ is the unique
rational in $I_\alpha$ with denominator at most $10^6$.
It does not establish that the physical coefficient lies
in this interval. The overlapping fitting windows can
share a bias from omitted sectors that is not measured
by their difference.

For the full-series comparisons, a divisor sieve supplies
the coefficients of the finite sum, whose ball enclosure
is enlarged by the tail bound~\eqref{sm:tailbound}.
At $N_f=1$, the evaluation uses $900$-digit ball arithmetic,
with the omitted tail smaller than $10^{-15}\eta$, where
$\eta$ is the certified critical-value allowance.
At $N_f=2,4$, it uses $950$ digits and a tail bound below
$10^{-910}$. A separate implementation sums the
polylogarithms and their derivatives in the Lambert
representation using $200$-digit point arithmetic and
a $185$-digit comparison threshold. This comparison
checks the evaluation of the proposed function, but
does not replace the ball enclosures.

Numerical inputs entering a subtraction remain decimal
strings until they are parsed by arbitrary-precision
arithmetic. Binary floating point is used only for final
plotting coordinates and predictor diagnostics.
The production environment is Python~3.10.13,
python-flint~0.9.0 with FLINT~3.6.0, and mpmath~1.3.0.
The least-squares fits use mpmath, while rational
reconstruction operates on the exported decimals with
exact integer and rational arithmetic.

\bibliographystyle{ytamsalpha}
\def\arxivfont{\rm}
\begingroup\small
\bibliography{references}\endgroup
\end{document}